\documentclass[12pt]{article}

\usepackage{amsmath,amssymb,epsfig,subfigure,color,soul,fleqn,rotating}
\usepackage{ulem,slashed}
\usepackage{cite}
\usepackage[colorlinks,citecolor=blue,urlcolor=blue,linkcolor=blue]{hyperref}
\usepackage{xcolor}

\newcommand{\be}{\begin{eqnarray}}
\newcommand{\ee}{\end{eqnarray}}
\newcommand{\nee}{\nonumber\end{eqnarray}}

\newcommand{\drbar}{{\overline{\rm DR}}}

\newcommand{\mch}[1] {m_{\ti \x^+_{#1}}}
\newcommand{\mnt}[1] {m_{\ti \x^0_{#1}}}

\newcommand{\msg}    {m_{\ti g}}
\newcommand{\msu}[1] {m_{\ti u_{#1}}}
\newcommand{\msd}[1] {m_{\ti d_{#1}}}

\def\gev             {{\rm GeV}}

\newcommand{\gsim}{\;\raisebox{-0.9ex}
           {$\textstyle\stackrel{\textstyle >}{\sim}$}\;}
\newcommand{\lsim}{\;\raisebox{-0.9ex}
          {$\textstyle\stackrel{\textstyle<}{\sim}$}\;}

\def\be            {\begin{equation}}
\def\ee            {\end{equation}}
\def\bea            {\begin{eqnarray}}
\def\eea            {\end{eqnarray}}

\definecolor{mybrown}{cmyk}{0,0.9,1.5,0.3}

\def\a              {\alpha}
\def\b               {\beta}

\def\g               {\gamma}

\def\x               {\chi}
\def\ti              {\tilde}
\def\sq              {\ti q}
\def\st              {\ti t}
\def\sc              {\ti c}
\def\sb              {\ti b}
\def\ch              {\ti \x^\pm}

\def\nt              {\ti \x^0}
\def\sg              {\ti g}
\def\stau            {\ti \tau}
\def\sneut           {\ti \nu}

\def\su                {\ti{u}}
\def\sto                  {\ti{t}}
\def \sca                 {\ti{c}}
\def\sd                {\ti{d}}
\def\ss                  {\ti{s}}
\def\sbo                 {\ti{b}}

\newcommand{\AddrGAKUGEI}{%
 \it Department of Physics, Tokyo Gakugei University, Koganei,
Tokyo 184-8501, Japan\\}

\newcommand{\AddrHEPHY}{%
 \it Institut f\"ur Hochenergiephysik der \"Osterreichischen Akademie
der Wissenschaften, A-1050 Vienna, Austria\\}

\newcommand{\AddrElena}{%
 \it VRVis Zentrum f\"ur Virtual Reality und Visualisierung Forschungs-GmbH,  A-1220 Vienna, Austria\\}

\newcommand{\AddrKEK}{%
 \it Institute of Particle and Nuclear Studies, High Energy Accelerator Research Organization (KEK), Ibaraki 305-0801, Japan\\}

\newcommand{\AddrAkimasa}{%
 \it The Graduate University for Advanced Studies (SOKENDAI), Hayama 240-0193, Japan\\}

\newcommand{\AddrIshikawa}{%
 \it International Center for Elementary Particle Physics, University of Tokyo, Tokyo 113-0033, Japan\\}

\title{\bf Imprint of SUSY in radiative $B$-meson decays}

\author{Helmut~Eberl${}^{1}$, Keisho~Hidaka${}^{2}$, Elena~Ginina${}^{1,3}$, Akimasa~Ishikawa${}^{4,5,6}$}

\date{
\small $^1$ \AddrHEPHY
       $^2$ \AddrGAKUGEI
       $^3$ \AddrElena
       $^4$ \AddrKEK
       $^5$ \AddrAkimasa
       $^6$ \AddrIshikawa
}

\definecolor{darkgreen}{rgb}{0,.5,0}

\begin{document}

\begin{flushright}
HEPHY-PUB 1024/21\\
KEK Preprint 2021-7\\
\end{flushright}

\begingroup
\let\newpage\relax
\maketitle
\endgroup

\maketitle
\thispagestyle{empty}

\begin{abstract}

	We study supersymmetric (SUSY) effects on $C_7(\mu_b)$ and $C'_7(\mu_b)$ 
	which are the Wilson coefficients (WCs) for $b \to s \gamma$ at b-quark 
	mass scale $\mu_b$ and are closely related to radiative $B$-meson decays.  
	The SUSY-loop contributions to $C_7(\mu_b)$ and $C'_7(\mu_b)$ are calculated 
	at leading order (LO) in the Minimal Supersymmetric Standard Model (MSSM) with general 
	quark-flavour violation (QFV). For the first time we perform a systematic MSSM 
	parameter scan for the WCs $C_7(\mu_b)$ and $C'_7(\mu_b)$ respecting all the relevant constraints, i.e. 
	the theoretical constraints from vacuum stability conditions and the experimental constraints, such as those 
	from $K$- and $B$-meson data and electroweak precision data, as well as recent limits on SUSY 
	particle masses and the 125 GeV Higgs boson data from LHC experiments.
	From the parameter scan we find the following: 
	(1) The MSSM contribution to Re($C_7(\mu_b)$) can be as large as $\sim \pm 0.05$, 
	  which could correspond to about 3$\sigma$ significance of New Physics (NP) signal 
	  in the future LHCb and Belle II experiments.
	(2) The MSSM contribution to Re($C'_7(\mu_b)$) can be as large as $\sim -0.08$, 
	  which could correspond to about 4$\sigma$ significance of NP signal 
	  in the future LHCb and Belle II experiments.
	(3) These large MSSM contributions to the WCs are mainly 
	  due to (i) large scharm-stop mixing and large scharm/stop involved 
	  trilinear couplings $T_{U23}$, $T_{U32}$ and $T_{U33}$, (ii) large 
	  sstrange-sbottom mixing and large sstrange-sbottom involved 
	  trilinear couplings $T_{D23}$, $T_{D32}$ and $T_{D33}$ and 
	  (iii) large bottom Yukawa coupling $Y_b$ for large $\tan\beta$ and large top Yukawa coupling $Y_t$.
	In case such large NP contributions to the WCs are really observed
	in the future experiments at Belle II and the LHCb Upgrade, 
    this could be the imprint of QFV SUSY 
	(the MSSM with general QFV) and would encourage to perform further studies of the 
	WCs $C'_7(\mu_b)$ and $C_7^{\rm MSSM}(\mu_b)$ at higher order (NLO/NNLO) level in this model. 
\end{abstract}

\clearpage

\section{Introduction}

Our present knowledge of elementary particle physics is very successfully described by 
the Standard Model (SM) of electroweak and strong interactions. 
This model has, however, several essential problems, such as naturalness and hierarchy problems. 
Moreover, it can not explain observed phenomena like the neutrino masses and mixings, the 
matter-antimatter asymmetry in our universe, and the origin of dark matter.
Hence, it is necessary to search for New Physics (NP) theory that solves these problems. 
The theory of Supersymmetry (SUSY) is still the most prominent candidate for such a NP 
theory solving the SM problems.

Here we study the influence of SUSY on $C_7(\mu_b)$ and $C'_7(\mu_b)$ which are the 
Wilson coefficients (WCs) for the quark flavour changing transition $b \to s \gamma$ at 
the b-quark mass scale $\mu_b$. They are closely related to radiative $B$-meson decays.
We calculate the SUSY-loop contributions to $C_7(\mu_b)$ and $C'_7(\mu_b)$ at leading 
order (LO) in the Minimal Supersymmetric Standard Model (MSSM) with general quark-flavour 
violation (QFV) due to squark generation mixing. In the numerical computation of the WCs, 
we perform a MSSM parameter scan respecting all the relevant theoretical and experimental 
constraints, such as those from vacuum stability conditions, those from $K$- and $B$-meson data, 
the 125 GeV Higgs boson data from LHC, and electroweak precision data, as well as recent 
limits on SUSY particle (sparticle) masses from LHC experiments.

On the experimental side, the WCs $C_7(\mu_b)$ and $C'_7(\mu_b)$ can be measured 
precisely in the ongoing and future experiments at Belle II and LHCb Upgrade 
~\cite{Belle-II_Physics_Book, LHCb-II_Physics_Book, Albrecht, Straub}.
There are many papers studying the radiative $B$-meson decays in the SM 
~\cite{Ali_92, Ali_93, Buras_93, Misiak_1996, Neubert, Buras, Misiak_2015, Misiak_2020}, 
the 2HDMs (Two-Higgs Doublet models) ~\cite{Borzumati_1998, Giudice, Soni_2000} and the MSSM 
~\cite{Okada, Hurth_2001, Kane, Lunghi_Porod, Lunghi}.

However, there is no systematic 
numerical study on the SUSY-loop contributions to $C_7(\mu_b)$ and $C'_7(\mu_b)$ 
{\it even at LO} in the MSSM with general QFV
\footnote{
To our knowledge, there is no complete next to leading order (NLO) computation of 
WCs $C_7(\mu_b)$ and $C'_7(\mu_b)$ in the MSSM with general QFV in the present literature. 
In ~\cite{Hurth_2011} gluino-squark loop contributions to the WCs $C_{7,8}(\mu_W)$ and 
$C'_{7,8}(\mu_W)$ at the weak scale $\mu_W$ are calculated at NLO of SUSY-QCD in the MSSM 
with general QFV. However, they did not perform a complete NLO computation of $C_7(\mu_b)$ 
and $C'_7(\mu_b)$.  
Here we remark that in ~\cite{Borzumati_1998, Giudice} the charged Higgs boson 
loop contributions to the WCs $C_{7,8}(\mu_W)$ and $C_7(\mu_b)$ are calculated 
at NLO of QCD in the 2HDMs.
}.
In this paper we thoroughly perform such a systematic study with special emphasis on the 
importance of SUSY QFV in order to clarify a possibility that an imprint of SUSY can be 
found in radiative $B$-meson decays, focusing on the WCs $C_7(\mu_b)$ and $C'_7(\mu_b)$. 

In the phenomenological study of the MSSM, usually quark-flavour conservation (QFC) is 
assumed, except for the quark-flavour violation stemming from the 
Cabibbo-Kobayashi-Maskawa (CKM) matrix. However, in general there can be SUSY QFV 
terms in the squark mass matrix. Especially important QFV terms are the mixing terms 
between the 2nd and the 3rd squark generations, such as $\sca_{L,R}-\st_{L,R}$ and 
$\ti{s}_{L,R}-\ti{b}_{L,R}$ mixing terms, where $\sca$, $\st$, $\ti{s}$ and $\ti{b}$ 
are the charm-, top-, strange- and bottom-squark, respectively.
In this study we put special emphasis on the influence of the SUSY QFV due to 
$\sca_{L,R}-\st_{L,R}$ and $\ti{s}_{L,R}-\ti{b}_{L,R}$ mixings on the WCs 
$C_7(\mu_b)$ and $C'_7(\mu_b)$.

In our analysis we assume that there is no SUSY lepton-flavour violation.
We also assume that R-parity is conserved and that the lightest neutralino $\nt_1$
is the lightest SUSY particle (LSP). We work in the MSSM with real parameters, 
except for the CKM matrix. 

In the following section we introduce the SUSY QFV parameters originating from 
the squark mass matrices. Details about our parameters scan are given in 
Section~\ref{sec:full scan}. In Section~\ref{sec:WC} we define the relevant WCs 
and analyze their behaviour in the MSSM with QFV. The conclusions are in 
Section~\ref{sec:concl}. All relevant constraints are listed in Appendix A.

%
\section{Squark mass matrices in the MSSM with flavour violation}
\label{sec:sq.matrix}
%
In the super-CKM basis of $\sq_{0 \gamma} =
(\sq_{1 {\rm L}}, \sq_{2 {\rm L}}, \sq_{3 {\rm L}}$,
$\sq_{1 {\rm R}}, \sq_{2 {\rm R}}, \sq_{3 {\rm R}}),~\gamma = 1,...6,$  
with $(q_1, q_2, q_3)=(u, c, t),$ $(d, s, b)$, the up-type and down-type squark mass matrices 
${\cal M}^2_{\tilde{q}},~\tilde{q}=\tilde{u},\tilde{d}$, at the SUSY scale have the following 
most general $3\times3$ block form~\cite{Allanach:2008qq}:
\begin{equation}
    {\cal M}^2_{\tilde{q}} = \left( \begin{array}{cc}
        {\cal M}^2_{\tilde{q},LL} & {\cal M}^2_{\tilde{q},LR} \\[2mm]
        {\cal M}^2_{\tilde{q},RL} & {\cal M}^2_{\tilde{q},RR} \end{array} \right), \quad \tilde{q}=\tilde{u},\tilde{d}\,.
 \label{EqMassMatrix1}
\end{equation}
Non-zero off-diagonal terms of the $3\times3$ blocks ${\cal M}^2_{\tilde{q},LL},~{\cal M}^2_{\tilde{q},RR},~{\cal M}^2_{\tilde{q},LR}$ 
and ${\cal M}^2_{\tilde{q},RL}$ in Eq.~(\ref{EqMassMatrix1}) explicitly break quark-flavour in the squark sector of the MSSM.
The left-left and right-right blocks in Eq.~(\ref{EqMassMatrix1}) are given by
\begin{eqnarray}
    & &{\cal M}^2_{\tilde{u}(\tilde{d}),LL} = M_{Q_{u(d)}}^2 + D_{\tilde{u}(\tilde{d}),LL}{\bf 1} + \hat{m}^2_{u(d)}, \nonumber \\
    & &{\cal M}^2_{\tilde{u}(\tilde{d}),RR} = M_{U(D)}^2 + D_{\tilde{u}(\tilde{d}),RR}{\bf 1} + \hat{m}^2_{u(d)},
     \label{EqM2LLRR}
\end{eqnarray}
where $M_{Q_{u}}^2=V_{\rm CKM} M_Q^2 V_{\rm CKM}^{\dag}$, $M_{Q_{d}}^2 \equiv M_Q^2$, 
$M_{Q,U,D}$ are the hermitian soft SUSY-breaking mass matrices of the squarks, 
$D_{\tilde{u}(\tilde{d}),LL}$, $D_{\tilde{u}(\tilde{d}),RR}$ are the $D$-terms, and  
$\hat{m}_{u(d)}$ are the diagonal mass matrices of the up(down)-type quarks.
$M_{Q_{u}}^2$ is related with $M_{Q_{d}}^2$
by the CKM matrix $V_{\rm CKM}$ due to the $SU(2)_{\rm L}$ symmetry.
The left-right and right-left blocks of Eq.~(\ref{EqMassMatrix1}) are given by
\begin{eqnarray}
 {\cal M}^2_{\tilde{u}(\tilde{d}),RL} = {\cal M}^{2\dag}_{\tilde{u}(\tilde{d}),LR} &=&
\frac{v_2(v_1)}{\sqrt{2}} T_{U(D)} - \mu^* \hat{m}_{u(d)}\cot\beta(\tan\beta),
\label{M2sqdef}
\end{eqnarray}
where $T_{U,D}$ are the soft SUSY-breaking trilinear 
coupling matrices of the up-type and down-type squarks entering the Lagrangian 
${\cal L}_{int} \supset -(T_{U\alpha \beta} \ti{u}^\dagger _{R\a}\ti{u}_{L\b}H^0_2 $ 
$+ T_{D\alpha \beta} \ti{d}^\dagger _{R\a}\ti{d}_{L\b}H^0_1)$,
$\mu$ is the higgsino mass parameter, and 
$\tan\beta = v_2/v_1$ with $v_{1,2}=\sqrt{2} \left\langle H^0_{1,2} \right\rangle$.
The squark mass matrices are diagonalized by the $6\times6$ unitary matrices $U^{\tilde{q}}$,
$\tilde{q}=\tilde{u},\tilde{d}$, such that
\begin{eqnarray}
&&U^{\tilde{q}} {\cal M}^2_{\tilde{q}} (U^{\tilde{q} })^{\dag} = {\rm diag}(m_{\tilde{q}_1}^2,\dots,m_{\tilde{q}_6}^2)\,,
\label{Umatr}
\end{eqnarray}
with $m_{\tilde{q}_1} < \dots < m_{\tilde{q}_6}$.
The physical mass eigenstates
$\sq_i, i=1,...,6$ are given by $\sq_i =  U^{\sq}_{i \alpha} \sq_{0\alpha} $.

In this paper we focus on the $\ti{c}_L - \ti{t}_L$, $\ti{c}_R - \ti{t}_R$, $\ti{c}_R - \ti{t}_L$, 
$\ti{c}_L - \ti{t}_R$, $\ti{s}_L - \ti{b}_L$, $\ti{s}_R - \ti{b}_R$, $\ti{s}_R - \ti{b}_L$, and 
$\ti{s}_L - \ti{b}_R$ mixing which is described by the QFV parameters $M^2_{Q_{u}23} \simeq M^2_{Q23}$, 
$M^2_{U23}$, $T_{U23}$, $T_{U32}$, $M^2_{Q23}$, $M^2_{D23}$, $T_{D23}$ and $T_{D32}$, respectively. 
We will also often refer to the QFC parameter $T_{U33}$ and $T_{D33}$ which induces the 
$\ti{t}_L - \ti{t}_R$ and $\ti{b}_L - \ti{b}_R$ mixing, respectively,  
and plays an important role in this study.\\
The slepton parameters are defined analogously to the squark ones. 
All the parameters in this study are assumed to be real, except the 
CKM matrix $V_{CKM}$.

\section{Parameter scan}
\label{sec:full scan}

In our MSSM-parameter scan we take into account theoretical constraints 
from vacuum stability conditions and experimental constraints from $K$- 
and $B$-meson data, the $H^0$ mass and coupling data and electroweak precision 
data, as well as limits on SUSY particle masses from recent LHC experiments 
(see Appendix A). Here $H^0$ is the discovered SM-like Higgs boson which we 
identify as the lightest $CP$ even neutral Higgs boson $h^0$ in the MSSM. 
Concerning squark generation mixings, we only consider the 
mixing between the second and third generation of squarks. The mixing between 
the first and the second generation squarks is strongly constrained by the 
$K$- and $D$-meson data ~\cite{Gabbiani:1996hi, PDG2020}. 
The experimental constraints on the mixing of the first and third generation squarks 
are not so strong \cite{Dedes}, but we don't consider this mixing since its 
effect is essentially similar to that of the mixing of the second and third 
generation squarks. We generate the input parameter points by using random numbers 
in the ranges shown in Table~\ref{table1}, where some parameters are fixed as given in 
the last box. All input parameters are $\drbar$ parameters defined at scale Q = 1 TeV, 
except $m_A(pole)$ which is the pole mass of the $CP$ odd Higgs boson $A^0$.
The parameters that are not shown explicitly are taken to be zero. 
The entire scan lies in the decoupling Higgs limit, i.e. in the scenarios 
with large $\tan\beta \geq 10$ and large $m_A \geq 1350$ GeV (see Table~\ref{table1}), respecting 
the fact that the discovered Higgs boson is SM-like. It is well known that the 
lightest MSSM Higgs boson $h^0$ is SM-like (including its couplings) in this limit.
We don't assume a GUT relation for the gaugino masses $M_1$, $M_2$, $M_3$.
\begin{table}[h!]
\footnotesize{
\caption{
Scanned ranges and fixed values of the MSSM parameters (in units of GeV or GeV$^2$, 
except for $\tan\beta$). The parameters that are not shown explicitly are 
taken to be zero. $M_{1,2,3}$ are the U(1), SU(2), SU(3) gaugino mass parameters.}
\begin{center}
\begin{tabular}{|c|c|c|c|c|c|}
    \hline
\vspace*{-0.3cm}
& & & & &\\
\vspace*{-0.3cm}
     $\tan\beta$ & $M_1$ &  $M_2$ & $M_3$ & $\mu$ &  $m_A(pole)$\\ 
& & & & &\\
    \hline
\vspace*{-0.3cm}
& & & & &\\
\vspace*{-0.3cm}
     10 $\div$ 80 & $100 \div 2500$ & $100 \div 2500$  & $2500 \div 5000$ & $100 \div 2500$ & $1350 \div 6000$\\
& & & & &\\
    \hline
    \hline
\vspace*{-0.3cm}
& & & & &\\
\vspace*{-0.3cm}
      $ M^2_{Q 22}$ & $ M^2_{Q 33}$ &  $|M^2_{Q 23}| $ & $ M^2_{U 22} $ & $ M^2_{U 33} $ &  $|M^2_{U 23}| $\\ 
& & & & &\\
     \hline
\vspace*{-0.3cm}
& & & & &\\
\vspace*{-0.3cm}
      $2500^2 \div 4000^2$ & $2500^2 \div 4000^2$ & $< 1000^2$  & $1000^2 \div 4000^2$ & $600^2 \div 3000^2$& $ < 2000^2$\\
& & & & &\\
    \hline
    \hline
\vspace*{-0.3cm}    
& & & & &\\
\vspace*{-0.3cm}      
      $ M^2_{D 22} $ & $ M^2_{D 33}$ &  $ |M^2_{D 23}|$ & $|T_{U 23}|  $ & $|T_{U 32}|  $ &  $|T_{U 33}|$\\ 
& & & & &\\
    \hline
\vspace*{-0.3cm}      
& & & & &\\
\vspace*{-0.3cm}  
       $ 2500^2 \div 4000^2$ & $1000^2 \div 3000^2 $ & $ < 2000^2$  & $< 4000 $ & $ < 4000$& $< 5000 $\\
& & & & &\\
 \hline 
\multicolumn{6}{c}{}\\[-3.6mm]  
\cline{1-4}
\vspace*{-0.3cm}      
     & & & \\
\vspace*{-0.3cm}      
     $ |T_{D 23}| $ & $|T_{D 32}|  $ &  $|T_{D 33}|$ &$|T_{E 33}| $\\ 
     & & & \\
    \cline{1-4}
\vspace*{-0.3cm}      
     & & & \\
\vspace*{-0.3cm}      
     $< 3000 $ & $< 3000 $& $ < 4000$& $ < 500$\\
     & & & \\
    \cline{1-4}
\end{tabular}\\[3mm]
\begin{tabular}{|c|c|c|c|c|c|c|c|c|}
    \hline
\vspace*{-0.3cm}      
    & & & & & & & &\\
\vspace*{-0.3cm}      
    $M^2_{Q 11}$ & $M^2_{U 11} $ &  $M^2_{D 11} $ & $M^2_{L 11}$ & $M^2_{L 22} $ &  $M^2_{L 33}$ & $M^2_{E 11}$&$M^2_{E 22}$ & $M^2_{E 33} $\\ 
    & & & & & & & &\\
    \hline
\vspace*{-0.3cm}      
    & & & & & & & &\\
\vspace*{-0.3cm}      
    $4500^2$ & $4500^2$ & $4500^2$  & $1500^2$ & $1500^2$ & $1500^2$& $1500^2$& $1500^2$&$1500^2$\\
    & & & & & & & &\\
    \hline
\end{tabular}
\end{center}
\label{table1}
}
\end{table}

All MSSM input parameters are taken as $\drbar$ parameters at the scale $Q = 1$~TeV, 
except $m_A(pole)$, and then are transformed by RGEs to those at the weak scale of 
$Q = \mu_W$ for the computation of the WCs $C_{7,8}(\mu_W)$ and $C'_{7,8}(\mu_W)$ in 
the MSSM. The masses and rotation matrices of the sfermions are renormalized at one-loop level 
by using the public code {\tt SPheno}-v3.3.8~\cite{SPheno1, SPheno2} based on the technique 
given in \cite{Pierce}.\\
%
\indent
From 8660000 input points generated in the scan 72904 points survived all constraints. 
These are 0.84\%. We show these survival points in all scatter plots in this article.

\section{WCs $\bold{C_7(\mu_b)}$ and $\bold{C'_7(\mu_b)}$ in the MSSM with QFV}
\label{sec:WC}

The effective Hamiltonian for the radiative transition $b \to s \gamma$ is given by 
\be
    H_{eff} = -{4G_F \over \sqrt 2} V_{tb} V^*_{ts} \sum_{i} C_i O_i,
    \label{eq_Heff}
\ee
where $G_F$ is the Fermi constant and $V_{tb} V^*_{ts}$ is a CKM factor.
The operators relevant to $b \to s \gamma$ are
\begin{eqnarray}
	O_2 & = & {\bar s_L} \gamma_\mu c_L {\bar c_L} \gamma^\mu b_L,               \nonumber \\
	O_7 & = & {e \over 16\pi^2} m_b {\bar s_L} \sigma^{\mu \nu} b_R F_{\mu \nu}, \nonumber \\
	O_8 & = & {g_s \over 16\pi^2} m_b {\bar s_L} \sigma^{\mu \nu} T^a b_R G^a_{\mu \nu},
	\label{eq_O}
\end{eqnarray}
and their chirality counterparts
\begin{eqnarray}
	O'_2 & = & {\bar s_R} \gamma_\mu c_R {\bar c_R} \gamma^\mu b_R,               \nonumber \\
	O'_7 & = & {e \over 16\pi^2} m_b {\bar s_R} \sigma^{\mu \nu} b_L F_{\mu \nu}, \nonumber \\
	O'_8 & = & {g_s \over 16\pi^2} m_b {\bar s_R} \sigma^{\mu \nu} T^a b_L G^a_{\mu \nu},
	\label{eq_O'}
\end{eqnarray}
where $m_b$ is the bottom quark mass, $e$ and $g_s$ are the electromagnetic and 
strong coupling, $F_{\mu \nu}$ and $G^a_{\mu \nu}$ the $U(1)_{em}$ and $SU(3)_c$ 
field-strength tensors, $T^a$ are colour generators, and the indices L,R denote 
the chirality of the quark fields. Here note that the SM contributions to $C'_{2,7,8}(\mu_W)$ 
are (almost) zero at LO.
The WCs $C_7(\mu_b)$ and $C'_7(\mu_b)$ at the bottom quark mass scale $\mu_b$ 
can be measured precisely in the experiments at Belle II and LHCb Upgrade 
~\cite{Belle-II_Physics_Book, LHCb-II_Physics_Book, Albrecht, Straub}.
We compute $C_7(\mu_b)$ and $C'_7(\mu_b)$ at LO in the MSSM with QFV and study 
the deviation of the MSSM predictions from their SM ones 
\footnote{
Here it is worth to mention that these WCs are related to the photon polarization in 
radiative $B$-meson decays. 
The helicity polarization of the external photon in $b \to s \gamma$ is defined as 
\be
P(b \to s \gamma) \equiv \frac{B(b \to s \gamma_R) - B(b \to s \gamma_L)}{B(b \to s \gamma_R) + B(b \to s \gamma_L)}.
   \label{P_gam1}	
\ee
At LO it is given as \cite{Kane}
\be
P(b \to s \gamma) = \frac{|C'_7(\mu_b)|^2 - |C_7(\mu_b)|^2}{|C'_7(\mu_b)|^2 + |C_7(\mu_b)|^2}.
   \label{P_gam2}	
\ee
In the SM $C'_7(\mu_b)$ is strongly suppressed by a factor $m_s/m_b$ and hence the photon in 
$b \to s \gamma$ decay is predominantly left-handed.
In principle, the photon polarization can be extracted from the measurement of radiative 
$B$-meson decays in the experiments such as Belle II and LHCb Upgrade 
\cite{Belle-II_Physics_Book, LHCb-II_Physics_Book, Choi, Grossman, Becirevic, LHCb_2015, 
Gronau_Soni, Belle_2006, BABAR_2008, Muheim, LHCb_2016, LHCb_2020}.
}
.
Following the standard procedure, first we compute $C_{7, 8}(\mu_W)$ and $C'_{7, 8}(\mu_W)$ 
at the weak scale $\mu_W$ at LO in the MSSM and then we compute $C_7(\mu_b)$ and $C'_7(\mu_b)$ 
by using the QCD RGEs for the scale evolution at leading log (LL) level \cite{Misiak_1996}
\footnote{
Here we comment on the RG running of the WCs at LL level.\\
\indent In footnote 5 of Ref.\cite{Kane}, it is argued as follows:\\
  In Ref.\cite{Borzumati} it has been pointed out that the gluino contribution to 
  the WCs $C_{7,8}^{(')}(\mu)$ is the sum of two different pieces, one proportional 
  to the gluino mass and one proportional to the bottom mass, which have a different 
  RG evolution (i.e. Eqs.(40) and (41) of \cite{Borzumati}, respectively). 
  However, it has been found that at LO this is equivalent to the usual SM 
  RG-evolution (i.e. Eqs.(13,14) of \cite{Kane} which correspond to Eq.(\ref{C7}) 
  of the present paper) once the running bottom mass $m_b(\mu_0)$ is used instead of 
  the pole mass $m_b$(pole) in the WCs $C^{(')}_i(\mu_0)$, where $\mu_0$ is the 
  high-energy matching scale (e.g. the electroweak scale $\mu_W$).\\
\indent We have also confirmed this point (fact) independently of Ref.\cite{Kane}.
Here, note that we have used the public code {\tt SPheno}-v3.3.8 \cite{SPheno1, SPheno2} 
in the computation of the WCs $C^{(')}_{7, 8}(\mu_0=160{\rm GeV})$, and that {\tt SPheno}-v3.3.8 
uses the running b-quark mass $m_b(\mu_0=160{\rm GeV})$ ({\it not} the pole mass $m_b$(pole)) 
in the computation of the $C^{(')}_{7, 8}(\mu_0=160{\rm GeV})$. 

Therefore, Eqs.(40) and (41) of \cite{Borzumati} are equivalent to the usual 
SM RG-evolution (i.e. Eq.(\ref{C7}) of the present paper) at LO. 

Moreover, just after Eq.(41) in Ref.\cite{Borzumati} it is clearly stated 
that the terms $R_{7b,\sg}(\mu_b)$ and $R_{8b,\sg}(\mu_b)$ turn out to be 
numerically very small with respect to the other terms on the right-hand 
sides of Eq.(41) for the RG running of the WCs. 
Here $R_{7b,\sg}(\mu_b)$ and $R_{8b,\sg}(\mu_b)$ are linear combinations 
of the WCs (such as $C^b_{i,\sg}(\mu_W)$ (i=15,16,19,20)) of the additional 
four-quark operators in Eq.(15) of \cite{Borzumati}, all of which are operators 
at NLO of QCD. Hence, the effects of the additional four-quark operators 
onto the RG running of $C_{7,8}^{(')}(\mu)$ are numerically very small.

Therefore, the contributions of the WCs of the new four-quark operators 
mentioned in \cite{Borzumati} (which are all at NLO of QCD) to the RG scale 
evolution (RG running) are numerically very small and hence the presence 
of the mentioned new four-quark operators can {\it not} change Eq.(\ref{C7}) 
in the present paper practically (essentially).
}
:
\begin{eqnarray}
	C_7(\mu_b) & = & \eta^{\frac{16}{23}}C_7(\mu_W) + \frac{8}{3}(\eta^{\frac{14}{23}}-\eta^{\frac{16}{23}})C_8(\mu_W) + \sum_{i=1}^{8} h_i\eta^{a_i}C_2(\mu_W)  \nonumber \\
	C'_7(\mu_b) & = & \eta^{\frac{16}{23}}C'_7(\mu_W) + \frac{8}{3}(\eta^{\frac{14}{23}}-\eta^{\frac{16}{23}})C'_8(\mu_W) + \sum_{i=1}^{8} h_i\eta^{a_i}C'_2(\mu_W), 
	    \label{C7}	
\end{eqnarray}	
where 
\begin{eqnarray}
	\eta & = & \alpha_s(\mu_W)/\alpha_s(\mu_b) \nonumber \\
	h_i & = & (\frac{626126}{272277}, -\frac{56281}{51730}, -\frac{3}{7}, -\frac{1}{14}, -0.6494, -0.0380, -0.0186, -0.0057) \nonumber \\
	a_i & = & (\frac{14}{23}, \frac{16}{23}, \frac{6}{23}, -\frac{12}{23}, 0.4086, -0.4230, -0.8994, 0.1456).
	    \label{eta}
\end{eqnarray}	

\noindent We take the NLO formula with 5 flavours for the strong coupling constant $\alpha_s(\mu)$ 
for $\mu_b \lesssim \mu \lesssim \mu_W$ \cite{Misiak_1996}:
\be
\alpha_s(\mu) = {\alpha_s(m_Z) \over v(\mu)} \left[1 - \frac{\beta_1}{\beta_0} \frac{\alpha_s(m_Z)}{4\pi} \frac{\ln v(\mu)}{v(\mu)} \right],
    \label{alpha_s}
\ee
\noindent where
\be
v(\mu) = 1 - \beta_0 \frac{\alpha_s(m_Z)}{2\pi} \ln(\frac{m_Z}{\mu}),
    \label{v_mu}
\ee
\noindent $\beta_0 = \frac{23}{3}$, $\beta_1 = \frac{116}{3}$ and $m_Z$ is the Z boson mass. 
We take $m_Z =91.2$ GeV and $\alpha_s(m_Z) = 0.1179$ \cite{PDG2020}.
The SM and MSSM contribution to $C_2(\mu_W) = C_2^{\rm SM}(\mu_W) + C_2^{\rm MSSM}(\mu_W)$ is 1 and 0 
at LO, respectively. 
The SM and MSSM contributions to $C^{'}_2(\mu_W) = C_2^{'\rm SM}(\mu_W) + C_2^{'\rm MSSM}(\mu_W)$ are 0 
at LO. In our numerical analysis, we take $\mu_W = 160 \rm GeV$ and $\mu_b = 4.8 \rm GeV$ \cite{Albrecht}.

\begin{figure*}[t!]
\centering

  { \mbox{\hspace*{0cm} \resizebox{7.0cm}{!}{\includegraphics{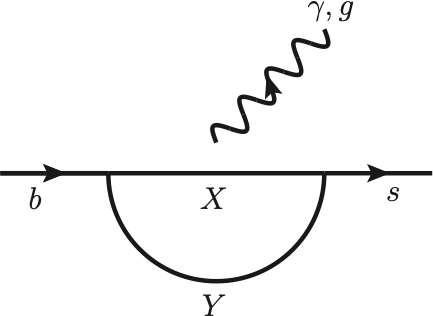}} \hspace*{0cm}}}
  \\

\caption{
The SM and MSSM one-loop contributions to the WCs $C_{7,8}(\mu_W)$ and $C'_{7,8}(\mu_W)$ 
at the weak scale $\mu_W$ for the transitions $b_R \to s_L \ \gamma_L, g_L$ and 
$b_L \to s_R \ \gamma_R, g_R$, respectively (see Eqs. (\ref{eq_Heff}, \ref{eq_O}, \ref{eq_O'})). 
Here $\gamma_L, g_L$ and $\gamma_R, g_R$ are left-handed photon, gluon and right-handed photon, 
gluon, respectively. The photon is emitted from any electrically charged line and the gluon 
from any colour charged line. 
For the SM one-loop contributions (X, Y) = (t/c/u, $W^+$). 
For the MSSM one-loop contributions (X, Y) = (stop/scharm/sup, chargino), 
(sbottom/sstrange/sdown, gluino), (sbottom/sstrange/sdown, neutralino) and (t/c/u, $H^+$), 
where stop/scharm/sup denotes top-, charm- , up-squark mixtures and so on.
}
\label{fig_b2sgamg_diagram}
\end{figure*}

We use the numerical results for $C^{(')}_{7,8}(\mu_W)$ at LO in the MSSM obtained from 
the public code {\tt SPheno}-v3.3.8~\cite{SPheno1, SPheno2}, which takes into account the 
following one-loop contributions to $C^{(')}_{7,8}(\mu_W)$ at the weak scale $\mu_W$ 
(see Fig. \ref{fig_b2sgamg_diagram}): \\
%

\noindent  1) SM one-loop contributions:\\
\indent up-type quark - $W^+$ loops\\

\noindent  2) MSSM one-loop contributions:\\
\indent up-type squark - chargino loops\\
\indent down-type squark - gluino loops\\
\indent down-type squark - neutralino loops\\
\indent up-type quark - $H^+$ loops\\

\noindent Here the chargino $\tilde\chi^{\pm}_{1,2}$ is a mixture of charged wino $\ti W^\pm$ and 
charged higgsino $\ti H^\pm$, the neutralino $\tilde\chi^0_{1,2,3,4}$ is a mixture of photino $\ti\g$, 
zino $\ti Z$ and two neutral higgsinos $\ti H^0_{1,2}$, and $H^+$ is the charged Higgs boson.

\begin{figure*}[t!]
\centering
 \subfigure []{
  { \mbox{\hspace*{0cm} \resizebox{5.8cm}{!}{\includegraphics{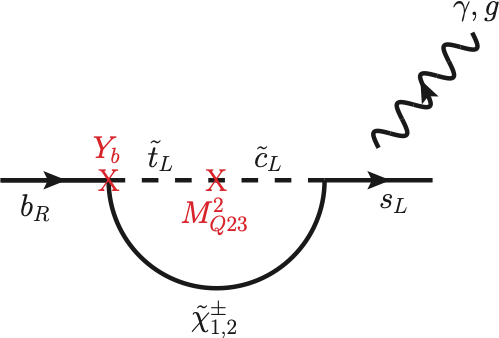}} \hspace*{0cm}}}
  \label{bR2sLgamg_chargino_loop_a}}
 \subfigure[]{
 { \mbox{\hspace*{0cm} \resizebox{6.1cm}{!}{\includegraphics{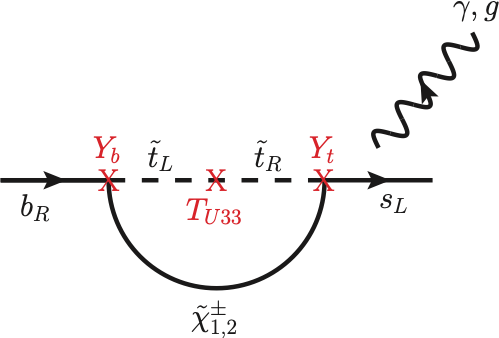}} \hspace*{0cm}}}
  \label{bR2sLgamg_chargino_loop_b}}
\caption{
Schematic illustration of important parts of the up-type squark - chargino loop 
contributions to $C_{7,8}(\mu_W)$ in terms of the mass-insertion approximation.
}
\label{bR2sLgamg_chargino_loop}
\end{figure*}
\begin{figure*}[t!]
\centering
 \subfigure []{
   { \mbox{\hspace*{0.0cm} \resizebox{6.0cm}{!}{\includegraphics{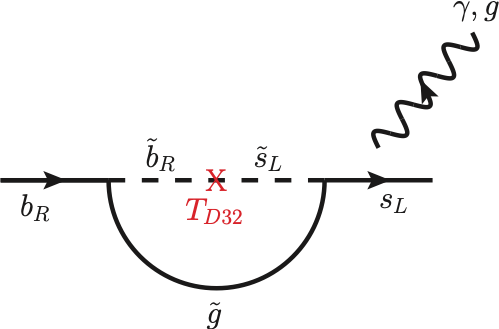}} \hspace*{0cm}}}
   \label{bR2sLgamg_sg_loop}}
 \subfigure[]{
   { \mbox{\hspace*{0cm} \resizebox{6.0cm}{!}{\includegraphics{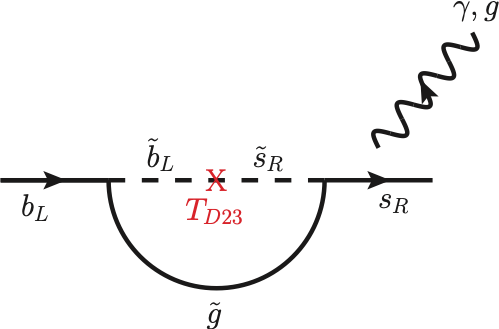}} \hspace*{0cm}}}
  \label{bL2sRgamg_sg_loop}}
\caption{
Schematic illustration of an important part of the down-type squark - gluino loop 
contributions to (a) $C_{7,8}(\mu_W)$ and (b) $C'_{7,8}(\mu_W)$ in terms of the 
mass-insertion approximation.
}
\label{b2sgamg_sg_loop}
\end{figure*}
\begin{figure*}[t!]
\centering
   {\mbox{\hspace*{0cm} \resizebox{6.2cm}{!}{\includegraphics{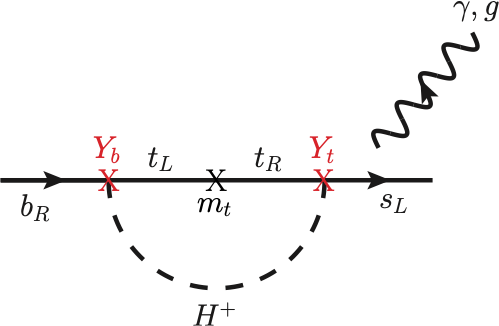}} \hspace*{0cm}}}
\caption{
Schematic illustration of the top quark - $H^+$ loop contribution to 
$C_{7,8}(\mu_W)$.
}
\label{bR2sLgamg_H_loop}
\end{figure*}

\noindent Before we show the results of the full parameter scan, we comment on 
the expected qualitative behavior of the MSSM one-loop contributions 
to $C^{(')}_7(\mu_b)$ at the bottom mass scale $\mu_b$.
We find that large squark trilinear couplings $T_{U23,32,33}$, 
$T_{D23,32,33}$, large $M^2_{Q23}$, $M^2_{U23}$, $M^2_{D23}$, large bottom 
Yukawa coupling $Y_b$ for large $\tan\beta$, and large top Yukawa coupling $Y_t$ 
can lead to large MSSM one-loop contributions to $C^{(')}_{7,8}(\mu_W)$ at the weak 
scale $\mu_W$, which results in large MSSM one-loop contributions to $C^{(')}_7(\mu_b)$ 
at the bottom mass scale $\mu_b$ (see Eq.(\ref{C7})).
%
%
This is mainly due to the following reasons:
\begin{itemize}

\item
The lighter up-type squarks $\su_{1,2,3}$ are strong $\sc_{L,R}$ - 
$\st_{L,R}$ mixtures for large $M^2_{Q23}$, 
$M^2_{U23}$, $T_{U23,32,33}$. The lighter down-type squarks $\sd_{1,2,3}$ 
are strong $\ss_{L,R}$ - $\sb_{L,R}$ mixtures 
for large $M^2_{Q23}$, $M^2_{D23}$, $T_{D23,32,33}$.
Here note that
$|T_{U23,32,33}|$ can be large due to large $Y_t$ (see Eqs.(\ref{eq:CCBfcU},\ref{eq:CCBfvU})) and that
$|T_{D23,32,33}|$ can be large due to large $Y_b$ for large $\tan\beta$ (see Eqs.(\ref{eq:CCBfcD},\ref{eq:CCBfvD})).
In the following we assume these setups.

\item
As for the up-type squark - chargino loop contributions to $C_7(\mu_W)$ 
and $C_8(\mu_W)$ which is the effective coupling for the transition 
$b_R \to s_L \, \gamma$ and $b_R \to s_L \, g$, respectively; \\
The $b_R$ - $\su_{1,2,3}$ - $\tilde\chi^{\pm}_{1,2}$ vertex which contains 
the $b_R$ - $\st_L$ - $\tilde H^{\pm}$ coupling can be enhanced by the large bottom 
Yukawa coupling $Y_b$ for large $\tan\beta$. 
%
%
The $s_L$ - $\su_{1,2,3}$ - $\tilde\chi^{\pm}_{1,2}$ vertex contains 
the $s_L$ - $\sc_L$ - $\tilde W^{\pm}$ coupling which is not CKM-suppressed
\footnote{
Note that the CKM-suppression factor $V^*_{ts}$ is factored out from WCs $C_i$ 
in their definition (see Eq.(\ref{eq_Heff})). Therefore, absence of the 
CKM-suppression factor in the one-loop diagram results in strong enhancement 
of the loop contribution to the WCs $C_i$. 
}. 
This vertex contains also the $s_L$ - $\st_R$ - $\tilde H^{\pm}$ coupling 
which is enhanced by the large top Yukawa coupling $Y_t$ despite the 
suppression due to the CKM factor $V^*_{ts}$. 
Hence, the up-type squark - chargino loop contributions 
to $C_{7,8}(\mu_W)$ can be enhanced by the large $Y_b$ for large $\tan\beta$ 
and the large $Y_t$, and further by the large $\sc_L$-$\st_L$ mixing term 
$M^2_{Q23}$ and the large $\st_L$-$\st_R$ mixing term $T_{U33}$ for which 
$\su_{1,2,3}$ contain a strong mixture of $\sc_L$, $\st_L$ and $\st_R$.
Important parts of this squark - chargino loop contributions to $C_{7,8}(\mu_W)$ 
are schematically illustrated in terms of the mass-insertion approximation 
in Fig. \ref{bR2sLgamg_chargino_loop}.

%


\item
As for the down-type squark - gluino loop contributions to $C_{7,8}(\mu_W)$; \\
The $b_R$ - $\sd_{1,2,3}$ - $\sg$ vertex which contains 
the $b_R$ - $\sb_R$ - $\sg$ coupling can be enhanced by the sizable QCD coupling. 
The $s_L$ - $\sd_{1,2,3}$ - $\sg$ vertex which contains 
the $s_L$ - $\ss_L$ - $\sg$ coupling can also be enhanced by the QCD coupling.
Furthermore, absence of the CKM-suppression factor in this loop diagram results 
in additional strong enhancement. 
Therefore, the down-type squark - gluino loop contributions to $C_{7,8}(\mu_W)$ 
can be enhanced by the sizable QCD coupling, and further 
by the large $\sb_R$-$\ss_L$ mixing term $T_{D32}$ for which $\sd_{1,2,3}$ contain 
a strong mixture of $\sb_R$ and $\ss_L$. Moreover, $|T_{D32}|$ can be large due to 
large $Y_b$ for large $\tan\beta$ (see Eq.(\ref{eq:CCBfvD})).
An important part of this squark - gluino loop contribution to $C_{7,8}(\mu_W)$ 
is schematically illustrated in terms of the mass-insertion approximation 
in Fig. \ref{bR2sLgamg_sg_loop}.


\item
As for the down-type squark - neutralino loop contributions to $C_{7,8}(\mu_W)$; \\
The $b_R$ - $\sd_{1,2,3}$ - $\tilde\chi^0_{1,2,3,4}$ vertex which contains 
the $b_R$ - $\sb_R$ - $\ti\g / \ti Z$ and $b_R$ - $\sb_L$ - $\ti H^0_1$ 
couplings with the latter coupling being proportional to $Y_b$ can be 
enhanced by large $Y_b$ for large $\tan\b$. 
The $s_L$ - $\sd_{1,2,3}$ - $\tilde\chi^0_{1,2,3,4}$ vertex contains 
the $s_L$ - $\ss_L$ - $\ti\g / \ti Z$ couplings. 
The absence of the CKM-suppression factor in this loop diagram results 
in additional strong enhancement.
Hence, the down-type squark - neutralino loop contributions to $C_{7,8}(\mu_W)$ 
can be enhanced by large $Y_b$ for large $\tan\b$, and further by 
the large $\sb_R$-$\ss_L$ and $\sb_L$-$\ss_L$ mixing terms ($T_{D32}$ and 
$M^2_{Q23}$), for which $\sd_{1,2,3}$ contain a strong mixture of 
$\sb_R$-$\ss_L$ and $\sb_L$-$\ss_L$. Moreover, $|T_{D32}|$ controlled by $Y_b$ 
can be large for large $tan\b$ (see Eq.(\ref{eq:CCBfvD})).

\item
As for the up-type quark - $H^+$ loop contributions to $C_{7,8}(\mu_W)$; \\
The $b_R$ - $t$ - $H^+$ vertex which contains 
the $b_R$ - $t_L$ - $H^+$ coupling can be enhanced by large $Y_b$ for large $\tan\b$. 
The $s_L$ - $t$ - $H^+$ vertex which contains 
the $s_L$ - $t_R$ - $H^+$ coupling can be enhanced by the large top-quark Yukawa 
coupling $Y_t$ despite the suppression due to the CKM factor $V^*_{ts}$.
Hence $t$ - $H^+$ loop contributions to $C_{7,8}(\mu_W)$ can be 
enhanced by large $Y_b$ for large $\tan\b$ and large $Y_t$.
The top quark - $H^+$ loop contribution to $C_{7,8}(\mu_W)$ is schematically 
illustrated in Fig. \ref{bR2sLgamg_H_loop}.

\item
As for the up-type squark - chargino loop contributions to $C'_7(\mu_W)$ 
and $C'_8(\mu_W)$ which are the effective couplings for the transition 
$b_L \to s_R \, \gamma$ and $b_L \to s_R \, g$, respectively; \\
From a similar argument one finds that these loop contributions to 
$C'_{7, 8}(\mu_W)$ should be small due to the very small s-quark Yukawa coupling $Y_s$.

\item
As for the down-type squark - gluino loop contributions to $C'_{7,8}(\mu_W)$; \\
The $b_L$ - $\sd_{1,2,3}$ - $\sg$ vertex which contains 
the $b_L$ - $\sb_L$ - $\sg$ coupling can be enhanced by the sizable QCD coupling. 
The $s_R$ - $\sd_{1,2,3}$ - $\sg$ vertex which contains 
the $s_R$ - $\ss_R$ - $\sg$ coupling can also be enhanced by the QCD coupling.
Absence of the CKM-suppression factor in this loop diagram results 
in additional strong enhancement.
Therefore, the down-type squark - gluino loop contributions to $C'_{7,8}(\mu_W)$ 
can be enhanced by the sizable QCD couplings, and further 
by large $\sb_L$-$\ss_R$ mixing term $T_{D23}$ for which $\sd_{1,2,3}$ contain 
a strong mixture of $\sb_L$ and $\ss_R$. Moreover, $|T_{D23}|$ can be large 
due to large $Y_b$ for large $\tan\beta$ (see Eq.(\ref{eq:CCBfvD})).
An important part of this squark - gluino loop contribution to $C'_{7,8}(\mu_W)$ 
is schematically illustrated in terms of the mass-insertion approximation in 
Fig. \ref{bL2sRgamg_sg_loop}.

\item
As for the down-type squark - neutralino loop contributions to $C'_{7,8}(\mu_W)$; \\
The $b_L$ - $\sd_{1,2,3}$ - $\tilde\chi^0_{1,2,3,4}$ vertex which contains 
the $b_L$ - $\sb_L$ - $\ti\g / \ti Z$ and $b_L$ - $\sb_R$ - $\ti H^0_1$ 
couplings with the latter coupling being proportional to $Y_b$ can be 
enhanced by large $Y_b$ for large $\tan\b$. 
The $s_R$ - $\sd_{1,2,3}$ - $\tilde\chi^0_{1,2,3,4}$ vertex contains 
the $s_R$ - $\ss_R$ - $\ti\g / \ti Z$ coupling.
Absence of the CKM-suppression factor in this loop diagram results 
in additional strong enhancement.
Hence, the down-type squark - neutralino loop contributions to 
$C'_{7,8}(\mu_W)$ can be enhanced by large $Y_b$ for large $\tan\b$, 
and further by large $\sb_L$-$\ss_R$ and $\sb_R$-$\ss_R$ 
mixing terms $T_{D23}$ and $M^2_{D23}$, for which $\sd_{1,2,3}$ 
contain strong mixtures of $\sb_L$-$\ss_R$ and $\sb_R$-$\ss_R$. 
Moreover, $|T_{D23}|$ controlled by $Y_b$ can be large for large 
$tan\b$ (see Eq.(\ref{eq:CCBfvD})).

\item
As for the up-type quark - $H^+$ loop contributions to $C'_{7,8}(\mu_W)$; \\
These contributions turn out to be very small due to the very small $Y_s$.

\end{itemize}

In the following we will show scatter plots in various planes obtained 
from the MSSM parameter scan described above (see Table \ref{table1}), 
respecting all the relevant constraints (see Appendix A).\\
\indent In Fig.~\ref{C7MSSM_C7'} we show scatter plots for $C^{\rm MSSM}_7(\mu_b)$ and $C'_7(\mu_b)$.
In Fig.~\ref{ReC7'_ImC7'} we show a scatter plot in the $Re(C'_7(\mu_b))$-$Im(C'_7(\mu_b))$ plane.
We see that the MSSM contribution to $Re(C'_7(\mu_b))$ can be as large 
as $\sim -0.07$, which could correspond to an about 4$\sigma$ New Physics (NP)         
signal significance in the combination of the future LHCb Upgrade (Phase III) 
and Belle II (Phase II) experiments (see Figure A.13 of \cite{Albrecht}). 
Note that $|Im(C'_7(\mu_b))|$ is very small (\lsim 0.004) and that $C'_7(\mu_b) \simeq 0$ in the SM.\\
%
%
In Fig.~\ref{ReC7MSSM_ImC7MSSM} we show the scatter plot in the 
$Re(C^{\rm MSSM}_7(\mu_b))$-$Im(C^{\rm MSSM}_7(\mu_b))$ plane.
We see that the MSSM contribution to $Re(C_7(\mu_b))$ can be as large 
as $\sim -0.05$, which could correspond to an about 3$\sigma$ NP signal significance     
in the combination of the future LHCb Upgrade (50 $fb^{-1}$) and Belle II 
(50 $ab^{-1}$) experiments (see Figure 8 of \cite{Albrecht}).
Note that $|Im(C^{\rm MSSM}_7(\mu_b))|$ is very small (\lsim 0.003) and that 
the MSSM contribution $C^{\rm MSSM}_7(\mu_b)$ can be quite sizable compared to 
$C^{\rm SM}_7(\mu_b) \simeq -0.325$.\\
%
%
In Fig.~\ref{ReC7MSSM_ReC7'} we show a scatter plot in the $Re(C^{\rm MSSM}_7(\mu_b))$-$Re(C'_7(\mu_b))$ 
plane. We see that the $Re(C'_7(\mu_b))$ and $Re(C^{\rm MSSM}_7(\mu_b))$ can be quite sizable 
simultaneously.\\ 

\indent Here we comment on the errors of the data on $C'_7(\mu_b)$ and $C^{NP}_7(\mu_b)$.  
The errors of the data on $C'_7(\mu_b)$ and $C^{NP}_7(\mu_b) \equiv C_7(\mu_b)-C^{\rm SM}_7(\mu_b)$ 
from the future $B$-meson experiments shown in Figure A.13 and Figure 8 of 
\cite{Albrecht} stem from experimental and theoretical errors. In general, $B$-meson observables 
are functions of the relevant WCs such as $C'_7(\mu_b)$ and $C_7(\mu_b)$. Hence, from the 
observed data on relevant $B$-meson observables one can determine (extract) the values of the WCs. 
The WCs thus determined (extracted) have two types of errors, one is the experimental 
error stemming from the (systematic and statistical) errors of the observable data 
and the other is the theoretical error due to the uncertainties of input parameters, 
such as the CKM matrix elements ($V_{ts}$, $V_{tb}$ ...), hadronic form factors and 
meson-decay constants, in the computation (prediction) of the observables 
by using the WCs (i.e. the effective couplings).\\
%
Here we remark the following points:
(i) As for the determination of $C_7(\mu_b)$ one can get much more precise 
information from the fully-inclusive $B(B \to X_s \gamma)$ measurement than 
from the measurement of the exclusive observables such as $B(B \to K^* \gamma)$  
\footnote{
Here note that $B(B \to X_s \gamma) \simeq B(b \to s \gamma)$ is proportional to 
$|C_7(\mu_b)|^2 + |C'_7(\mu_b)|^2$ at LO.
}
since the theoretical predictions for the exclusive observables involve 
hadronic form factors which have large theoretical uncertainties.
(ii) The fully-inclusive observable $B(B \to X_s \gamma)$ can be measured 
reliably and precisely at Belle II ~\cite{Belle-II_Physics_Book} whereas its 
measurement is very difficult at LHCb ~\cite{LHCb-II_Physics_Book}.
(iii) As a result, Belle II plays a specially important role in the 
precise determination (extraction) of $C_7(\mu_b)$ in the near future.\\
As for the experimental errors of the WCs $C'_7(\mu_b)$ and 
$C^{NP}_7(\mu_b)$ obtained (extracted) from the future $B$-meson experiments, 
Belle II is now planning to upgrade to accumulate about 5 times larger data (up to $\sim 250 \, ab^{-1}$)
\cite{Browder@Snowmass}. If this is realized, the (statistical) uncertainty 
of the observable data from Belle II could be reduced by a factor of about $\sqrt 5$.\\
As for the theoretical errors of the WCs $C'_7(\mu_b)$ and 
$C^{NP}_7(\mu_b)$ obtained (extracted) from the $B$-meson experiments, there is a sign 
of promising possibility of significant reduction of the theoretical errors in the future: 
Very recently M. Misiak et al. performed a new computation of $B(B \to X_s \gamma)$ 
in the SM at the NNLO in QCD \cite{Misiak_2020}. 
Taking into account the recently improved estimates of non-perturbative contributions, 
they have obtained $B(B \to X_s \gamma) = (3.40 \pm 0.17) \cdot 10^{-4}$ for $E_\gamma > 1.6 GeV$.
Compared with their previous SM prediction $B(B \to X_s \gamma) = (3.36 \pm 0.23) \cdot 10^{-4}$ 
\cite{Misiak_2015}, the theoretical uncertainty is now reduced from 6.8\% to 5.0\%.  
Note here that the Figure A.13 and Figure 8 of \cite{Albrecht} showing expected errors 
of $C'_7(\mu_b)$ and $C^{NP}_7(\mu_b)$ obtained (extracted) 
from the future $B$-meson experiments were made in the year 2017.\\
Hence, in case the significant reduction of the experimental and theoretical errors 
is achieved in the future, the NP signal significances for $Re(C'_7(\mu_b))$ and 
$Re(C_7^{NP}(\mu_b))$ in the MSSM could be significantly higher than those mentioned 
above which are 
about 4 $\sigma$ NP significances for $Re(C'_7(\mu_b))$ and 
about 3 $\sigma$ significance for $Re(C_7^{\rm MSSM}(\mu_b))$.\\
Thus, it is very important to improve the precision of both theory and experiment on $B$-meson physics 
by a factor about 1.5 or so in view of NP search (such as SUSY search).
Therefore, we strongly encourage theorists and experimentalists to challenge this task.\\
%
%

%
\begin{figure*}[t!]
\centering
 \subfigure []{
   { \mbox{\hspace*{-1cm} \resizebox{7.5cm}{!}{\includegraphics{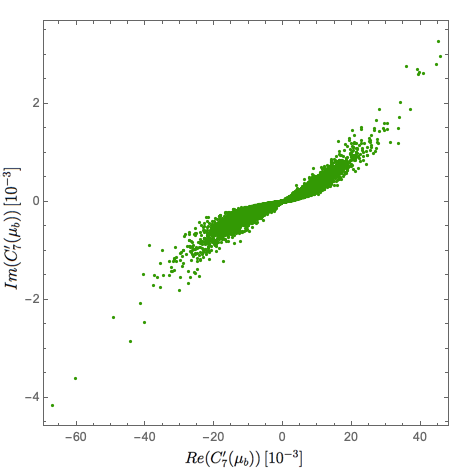}} \hspace*{0cm}}}
   \label{ReC7'_ImC7'}}
 \subfigure[]{
   { \mbox{\hspace*{0cm} \resizebox{7.5cm}{!}{\includegraphics{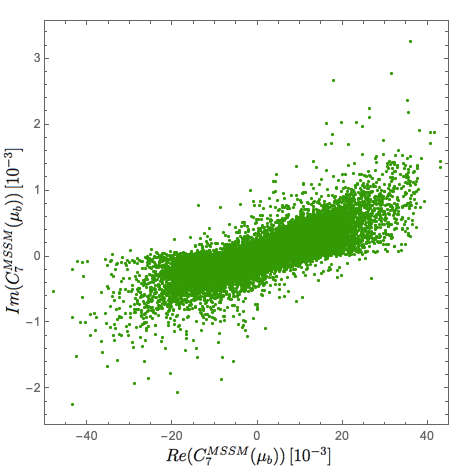}} \hspace*{0cm}}}
  \label{ReC7MSSM_ImC7MSSM}}\\
 \subfigure[]{
   { \mbox{\hspace*{0cm} \resizebox{7.5cm}{!}{\includegraphics{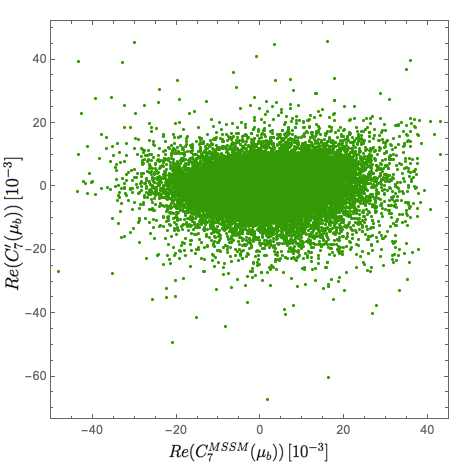}} \hspace*{0cm}}}
  \label{ReC7MSSM_ReC7'}}
\caption{
The scatter plot of the scanned parameter points within the ranges given in Table \ref{table1}
in the planes of (a) $Re(C'_7(\mu_b))$ - $Im(C'_7(\mu_b))$, (b) $Re(C^{\rm MSSM}_7(\mu_b))$ - $Im(C^{\rm MSSM}_7(\mu_b))$, 
and (c) $Re(C^{\rm MSSM} _7(\mu_b))$ - $Re(C'_7(\mu_b))$. 
}
\label{C7MSSM_C7'}
\end{figure*}

\begin{figure*}[t!]
\centering
\subfigure[]{
      { \mbox{\hspace*{-1cm} \resizebox{7.5cm}{!}{\includegraphics{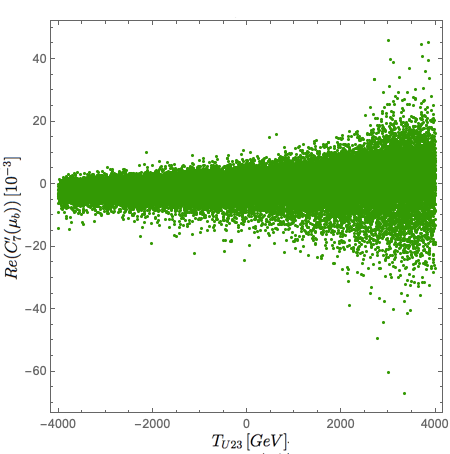}} \hspace*{0cm}}}
  \label{TU23_C'7}}
 \subfigure[]{
   { \mbox{\hspace*{0cm} \resizebox{7.5cm}{!}{\includegraphics{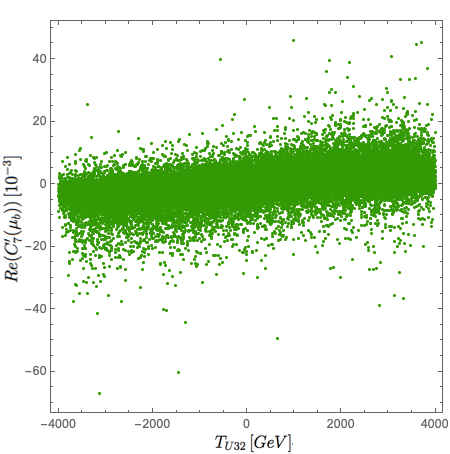}} \hspace*{0cm}}}
   \label{TU32_C'7}}\\
\subfigure[]{
      { \mbox{\hspace*{-1cm} \resizebox{7.5cm}{!}{\includegraphics{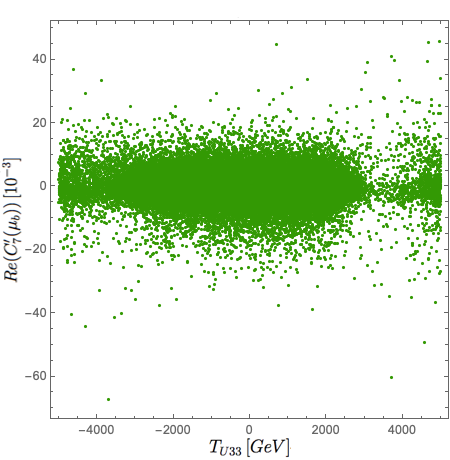}} \hspace*{0cm}}}
  \label{TU33_C'7}}
 \subfigure[]{
   { \mbox{\hspace*{0cm} \resizebox{7.5cm}{!}{\includegraphics{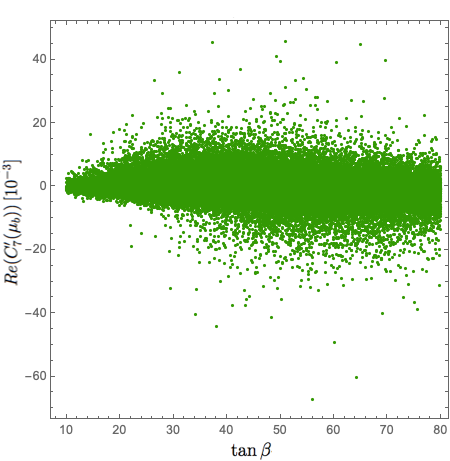}} \hspace*{0cm}}}
 \label{tanb_C'7}}
\caption{
The scatter plots of the scanned parameter points within the ranges given in Table \ref{table1} in the 
planes of (a) $T_{U23}$-$Re(C'_7(\mu_b))$, (b) $T_{U32}$-$Re(C'_7(\mu_b))$, (c) $T_{U33}$-$Re(C'_7(\mu_b))$ 
and (d) $\tan\b$-$Re(C'_7(\mu_b))$.
}
\label{TU23_tanb_C'7}
\end{figure*}

\begin{figure*}[t!]
\centering
\subfigure[]{
      { \mbox{\hspace*{-1cm} \resizebox{7.5cm}{!}{\includegraphics{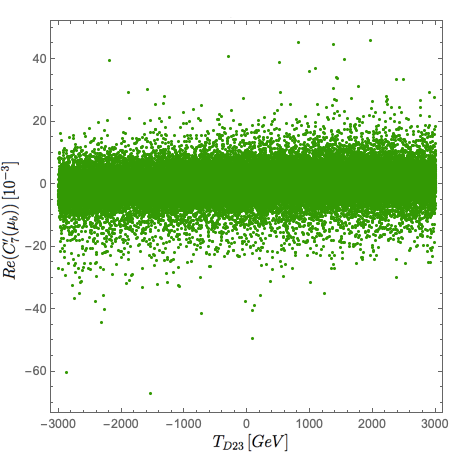}} \hspace*{0cm}}}
  \label{TD23_C'7}}
 \subfigure[]{
   { \mbox{\hspace*{0cm} \resizebox{7.5cm}{!}{\includegraphics{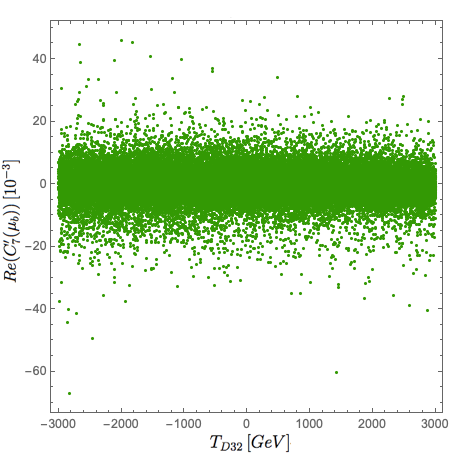}} \hspace*{0cm}}}
   \label{TD32_C'7}}\\
\subfigure[]{
      { \mbox{\hspace*{0cm} \resizebox{7.5cm}{!}{\includegraphics{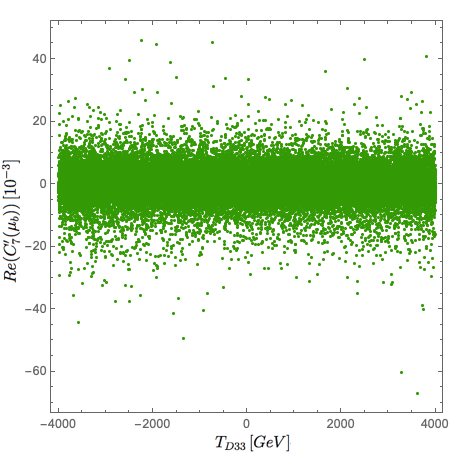}} \hspace*{0cm}}}
  \label{TD33_C'7}}
\caption{
The scatter plots of the scanned parameter points within the ranges given in Table \ref{table1} in the planes of 
(a) $T_{D23}$-$Re(C'_7(\mu_b))$, (b) $T_{D32}$-$Re(C'_7(\mu_b))$ and (c) $T_{D33}$-$Re(C'_7(\mu_b))$.
}
\label{TD_C'7}
\end{figure*}

\begin{figure*}[t!]
\centering
 \subfigure[]{
   { \mbox{\hspace*{-0.5cm} \resizebox{7.5cm}{!}{\includegraphics{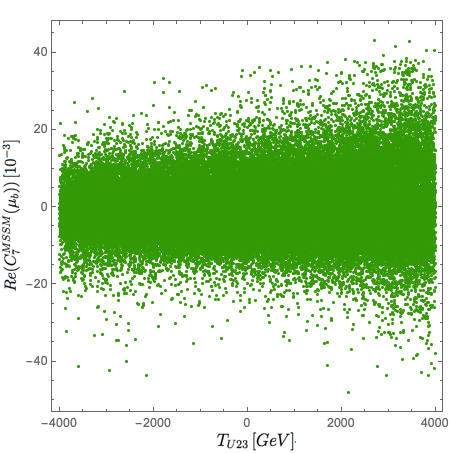}} \hspace*{0cm}}}
   \label{TU23_C7MSSM}}
 \subfigure[]{
   { \mbox{\hspace*{0cm} \resizebox{7.5cm}{!}{\includegraphics{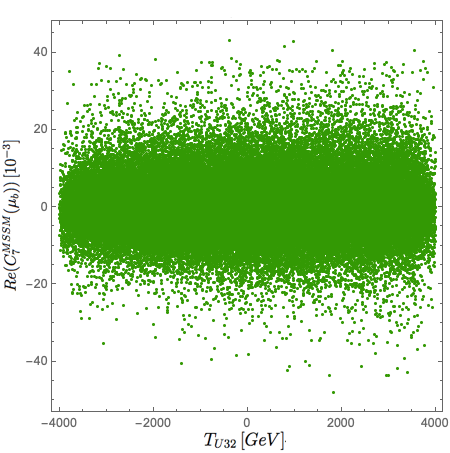}} \hspace*{0cm}}}
   \label{TU32_C7MSSM}}\\
 \subfigure[]{
   { \mbox{\hspace*{-0.5cm} \resizebox{7.5cm}{!}{\includegraphics{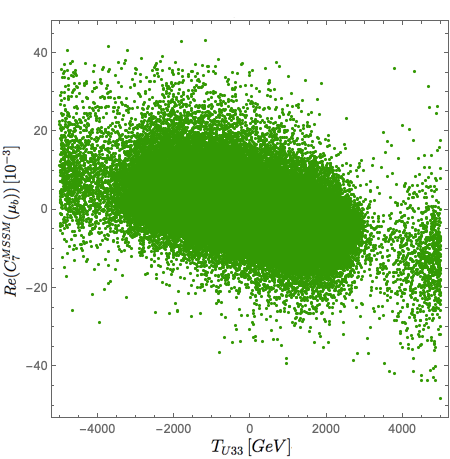}} \hspace*{0cm}}}
  \label{TU33_C7MSSM}}
 \subfigure[]{
   { \mbox{\hspace*{0cm} \resizebox{7.5cm}{!}{\includegraphics{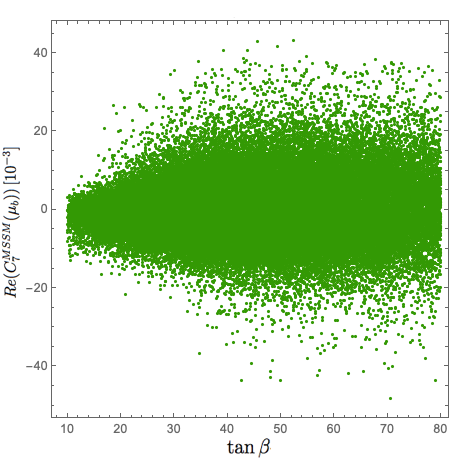}} \hspace*{0cm}}}
   \label{tanb_C7MSSM}}
\caption{
The scatter plot of the scanned parameter points within the ranges given in Table \ref{table1} 
in the planes of (a) $T_{U23}$-$Re(C^{\rm MSSM}_7(\mu_b))$, 
(b) $T_{U32}$-$Re(C^{\rm MSSM}_7(\mu_b))$, (c) $T_{U33}$-$Re(C^{\rm MSSM}_7(\mu_b))$ and (d) $\tan\b$-$Re(C^{\rm MSSM}_7(\mu_b))$.
}
\label{TU23_tanb_C7MSSM}
\end{figure*}

\begin{figure*}[t!]
\centering
  { \mbox{\hspace*{0cm} \resizebox{7.5cm}{!}{\includegraphics{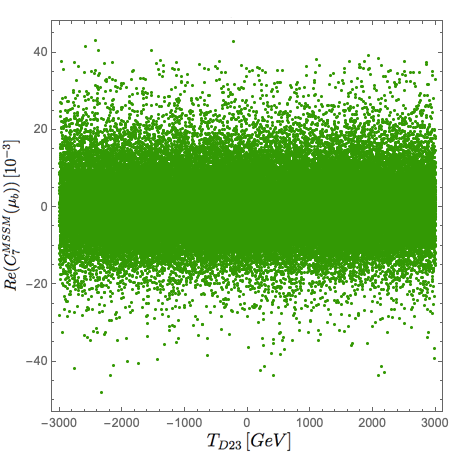}} \hspace*{0cm}}}
\caption{
The scatter plot of the scanned parameter points within the ranges given in Table \ref{table1} 
in the $T_{D23}$-$Re(C^{\rm MSSM}_7(\mu_b))$ plane.
}
\label{TD_C7MSSM}
\end{figure*}

\indent In Fig.~\ref{TU23_tanb_C'7} we show scatter plots in the $T_{U23}$-$Re(C'_7(\mu_b))$, 
$T_{U32}$-$Re(C'_7(\mu_b))$, $T_{U33}$-$Re(C'_7(\mu_b))$ and $\tan\b$-$Re(C'_7(\mu_b))$ planes.
From Fig.~\ref{TU23_C'7} we see that $Re(C'_7(\mu_b)) \simeq Re(C_7^{'\rm MSSM}(\mu_b))$ can be sizable 
($-0.07 \lsim Re(C'_7(\mu_b)) \lsim 0.05$) for large $T_{U23}$ ($\gsim$ 3 TeV).
From Fig.~\ref{TU32_C'7} we see that $Re(C'_7(\mu_b))$ can be large for large $|T_{U32}|$:
$-0.07 \lsim Re(C'_7(\mu_b)) \lsim 0.025$ for $T_{U32} \lsim-2$ TeV and 
$-0.04 \lsim Re(C'_7(\mu_b)) \lsim 0.045$ for $T_{U32} \gsim 2$ TeV.
A significant correlation between $Re(C'_7(\mu_b))$ and $T_{U32}$ can be seen.
From Fig.~\ref{TU33_C'7} we see that $Re(C'_7(\mu_b))$ can be large for large $|T_{U33}| \gsim 3$ TeV. 
The fewer scatter points around $T_{U33} = 3.5$ TeV is due to the fact that the $m_{h^0}$ bound 
tends to be violated around this point. 
From Fig.~\ref{tanb_C'7} we see that $Re(C'_7(\mu_b))$ can be large for large $\tan\b$:
$-0.07 \lsim Re(C'_7(\mu_b)) \lsim 0.05$ for $\tan\b \gsim 40$. 
All of these features are consistent with our expectation from the argument above.\\
%
%
\indent In Fig.~\ref{TD_C'7} we show scatter plots in the $T_{D23}$-$Re(C'_7(\mu_b))$, 
$T_{D32}$-$Re(C'_7(\mu_b))$ and $T_{D33}$-$Re(C'_7(\mu_b))$ planes.
From Fig.~\ref{TD23_C'7} and Fig.~\ref{TD32_C'7} we see that $Re(C'_7(\mu_b)) \simeq Re(C_7^{'\rm MSSM}(\mu_b))$ 
can be large ($-0.07 \lsim Re(C'_7(\mu_b)) \lsim 0.05$) for large $T_{D23}$, $T_{D32}$ ($\lsim -1.5$ TeV). 
An appreciable correlation between $T_{D23}$ and $Re(C'_7(\mu_b))$ can be seen in Fig.~\ref{TD23_C'7}.
From Fig.~\ref{TD33_C'7} we see that it can be large for large $|T_{D33}| \gsim 2$ TeV. 
These behaviors are also consistent with our expectation.\\
%
\indent 
In Fig.~\ref{TU23_tanb_C7MSSM} we show scatter plots in the planes of 
$T_{U23}$-$Re(C^{\rm MSSM}_7(\mu_b))$, $T_{U32}$-$Re(C^{\rm MSSM}_7\\
(\mu_b))$, $T_{U33}$-$Re(C^{\rm MSSM}_7(\mu_b))$ and $\tan\b$-$Re(C^{\rm MSSM}_7(\mu_b))$.
From Fig.~\ref{TU23_C7MSSM} and Fig.~\ref{TU32_C7MSSM} we see that $Re(C_7^{\rm MSSM}(\mu_b))$ 
can be sizable (up to $\sim \pm 0.05$) compared with $Re(C_7^{\rm SM}(\mu_b)) \simeq -0.325$ 
for large $T_{U23}$ and $T_{U32} ~(\gsim 2$ TeV).
%
%
From Fig.~\ref{TU33_C7MSSM} we see that it can be large for large $|T_{U33}|$:
$-0.03 \lsim Re(C_7^{\rm MSSM}(\mu_b)) \lsim 0.045$ for $T_{U33} \lsim -2$ TeV and 
$-0.05 \lsim Re(C_7^{\rm MSSM}(\mu_b)) \lsim 0.035$ for $T_{U33} \gsim 2$ TeV.
There is a significant correlation between $Re(C_7^{\rm MSSM}(\mu_b))$ and $T_{U33}$, 
which can be explained partly by the important contribution of 
Fig.~\ref{bR2sLgamg_chargino_loop_b} (see Eq.(\ref{C7})).
The fewer scatter points around $T_{U33} = 3.5$ TeV is again due to the fact that 
the $m_{h^0}$ bound tends to be violated around this point.
From Fig.~\ref{tanb_C7MSSM} we see that it can be large 
(up to $\sim \pm 0.05$) for large $\tan\b$ ($\gsim 40$).
These behaviors are also consistent with our expectation.\\
%
%
\indent In Fig.~\ref{TD_C7MSSM} we show scatter plots in the $T_{D23}$-$Re(C^{\rm MSSM}_7(\mu_b))$ plane.
We see $Re(C^{\rm MSSM}_7(\mu_b))$ can be sizable (up to $\sim \pm 0.05$) for any values of $T_{D23}$.
We have found that scatter plots in the $T_{D32}$-$Re(C^{\rm MSSM}_7(\mu_b))$ and 
$T_{D33}$-$Re(C^{\rm MSSM}_7(\mu_b))$ planes have similar behavior to that in the 
$T_{D23}$-$Re(C^{\rm MSSM}_7(\mu_b))$ plane.\\
%
%

%
%

%
%

%

\newpage
In order to see the relevant parameter dependences of $Re(C^{\rm MSSM}_7(\mu_b))$ and 
$Re(C'_7(\mu_b))$ in more detail, we take a reference scenario P1 where we have 
sizable $Re(C^{\rm MSSM}_7(\mu_b))$ and $Re(C'_7(\mu_b))$ and then variate the 
relevant parameters around this point P1. All MSSM input parameters for P1 
are shown in Table~\ref{table2}, where one has $(Re(C_7(\mu_b)), Im(C_7(\mu_b)) = 
(-0.370, -5.13\cdot 10^{-4})$, $(Re(C^{\rm SM}_7(\mu_b)), Im(C^{\rm SM}_7(\mu_b)) = 
(-0.325, 5.63\cdot 10^{-7})$, \\
$(Re(C^{\rm MSSM}_7(\mu_b)), Im(C^{\rm MSSM}_7(\mu_b)) = 
(-0.0441, -5.13\cdot 10^{-4})$ and
$(Re(C'_7(\mu_b)), \\Im(C'_7(\mu_b)) = 
(-0.0472, -1.81\cdot 10^{-3})$ with $C_7(\mu_b) = C^{\rm SM}_7(\mu_b) + C^{\rm MSSM}_7(\mu_b)$.\\
The scenario P1 satisfies all present experimental and theoretical constraints, see Appendix A. 
The resulting physical masses of the particles are shown in Table~\ref{physmasses}. 
The flavour decompositions of the lighter squarks $\su_{1,2,3}$ and $\sd_{1,2,3}$ are shown in Table~\ref{flavourdecomp}. 
For the calculation of the masses and the mixing, as well as for the low-energy observables, especially 
those in the B and K meson sectors (see Table~\ref{TabConstraints}), we use the public code 
{\tt SPheno} v3.3.8~\cite{SPheno1, SPheno2}. 
For the calculation of the coupling modifier $\kappa_b = C(h^0 b \bar{b})/C(h^0 b \bar{b})_{\rm SM}$ 
(or equivalently the deviation $DEV(b) \equiv \Gamma(h^0 \to b \bar{b})/\Gamma(h^0 \to b \bar{b})_{\rm SM} -1 ~
(= \kappa_b^2 -1)$ of the width $\Gamma(h^0 \to b \bar{b})$ from 
its SM value) we compute the width $\Gamma(h^0 \to b \bar{b})$ at full one-loop level in the MSSM with QFV 
by using the code developed by us \cite{Eberl:h2bb}. 
For the coupling modifier $\kappa_x = C(h^0 x x)/C(h^0 x x)_{\rm SM}$ with $x = g ~\mbox{or} ~\gamma$ (or the deviation 
$DEV(x) \equiv \Gamma(h^0 \to x x)/\Gamma(h^0 \to x x)_{\rm SM} -1 ~
(= \kappa_x^2 -1)$) we compute the width $\Gamma(h^0 \to x x)$ according to \cite{h2gagagg}.
We obtain $\kappa_b = 1.03$ (or $DEV(b) = 0.0686$), $\kappa_g = 0.994$ (or $DEV(g) = -0.0120$) and
$\kappa_\gamma = 1.0018$ (or $DEV(\gamma) = 0.0036$), which satisfy the LHC data in Table~\ref{TabConstraints}. 
For the B and K meson observables we get;
$B(b \to s \gamma) =  3.764 \cdot 10^{-4}$, $B(b \to s \ l^+ l^-) = 1.589 \cdot 10^{-6}$, 
$B(B_s \to \mu^+ \mu^-) = 2.5930 \cdot 10^{-9}$, $B(B^+ \to \tau^+ \nu) = 9.942 \cdot 10^{-5}$, 
$\Delta M_{B_s} = 17.180 [ps^{-1}]$, $|\epsilon_K| = 2.201 \cdot 10^{-3}$, $\Delta M_K = 2.304 \cdot 10^{-15} \ (GeV)$, 
$B(K^0_L \to \pi^0 \nu \bar{\nu}) = 2.295 \cdot 10^{-11}$, and $B(K^+ \to \pi^+ \nu \bar{\nu}) = 7.771 \cdot 10^{-11}$, 
all of which satisfy the constraints of Table~\ref{TabConstraints}.\\

%
\begin{table}[h!]
\footnotesize{
\caption{
The MSSM parameters for the reference point P1 (in units 
of GeV or GeV$^2$ expect for $\tan\beta$). All parameters are 
defined at scale Q = 1 TeV, except $m_A(pole)$. 
The parameters that are not shown here are taken to be zero.
}
\begin{center}
\begin{tabular}{|c|c|c|c|c|c|}
    \hline
\vspace*{-0.3cm}
& & & & &\\
\vspace*{-0.3cm}
     $\tan\beta$ & $M_1$ &  $M_2$ & $M_3$ & $\mu$ &  $m_A(pole)$\\ 
& & & & &\\
    \hline
\vspace*{-0.3cm}
& & & & &\\
\vspace*{-0.3cm}
     70 & 910 & 1970 & 2795 & 800 & 4970\\
& & & & &\\
    \hline
    \hline
\vspace*{-0.3cm}
& & & & &\\
\vspace*{-0.3cm}
      $M^2_{Q 22}$ & $ M^2_{Q 33}$ &  $M^2_{Q 23}$ & $ M^2_{U 22} $ & $ M^2_{U 33} $ &  $M^2_{U 23} $\\ 
& & & & &\\
     \hline
\vspace*{-0.3cm}
& & & & &\\
\vspace*{-0.3cm}
     3630$^2$ & 3365 $^2$ & -740$^2$ & 2755$^2$ & 1510 $^2$ & -1705$^2$\\
& & & & &\\
    \hline
    \hline
\vspace*{-0.3cm}    
& & & & &\\
\vspace*{-0.3cm}      
      $ M^2_{D 22} $ & $ M^2_{D 33}$ &  $ M^2_{D 23}$ & $T_{U 23}  $ & $T_{U 32}  $ &  $T_{U 33}$\\ 
& & & & &\\
    \hline
\vspace*{-0.3cm}      
& & & & &\\
\vspace*{-0.3cm}  
      2985$^2$ & 1270$^2$ & -1820$^2$ & 2700 & -260 & 4995\\
& & & & &\\
 \hline 
\multicolumn{6}{c}{}\\[-3.6mm]  
\cline{1-4}
\vspace*{-0.3cm}      
     & & & \\
\vspace*{-0.3cm}      
     $ T_{D 23} $ & $T_{D 32}  $ &  $ T_{D 33}$ &$T_{E 33} $\\ 
     & & & \\
    \cline{1-4}
\vspace*{-0.3cm}      
     & & & \\
\vspace*{-0.3cm}      
     -2330 & -335 & 3675 & -335\\
     & & & \\
    \cline{1-4}
\end{tabular}\\[3mm]
\begin{tabular}{|c|c|c|c|c|c|c|c|c|}
    \hline
\vspace*{-0.3cm}      
    & & & & & & & &\\
\vspace*{-0.3cm}      
    $M^2_{Q 11}$ & $M^2_{U 11}$ &  $M^2_{D 11}$ & $M^2_{L 11}$ & $M^2_{L 22}$ & $M^2_{L 33}$ & $M^2_{E 11}$ & $M^2_{E 22}$ & $M^2_{E 33} $\\ 
    & & & & & & & &\\
    \hline
\vspace*{-0.3cm}      
    & & & & & & & &\\
\vspace*{-0.3cm}      
    $4500^2$ & $4500^2$ & $4500^2$  & $1500^2$ & $1500^2$ & $1500^2$ & $1500^2$ & $1500^2$ & $1500^2$\\
    & & & & & & & &\\
    \hline
\end{tabular}
\end{center}
\label{table2}
}
\end{table}

\begin{table}
\caption{Physical masses in GeV of the particles for the scenario of Table~\ref{table2}.}
\begin{center}
\begin{tabular}{|c|c|c|c|c|c|}
  \hline
  $\mnt{1}$ & $\mnt{2}$ & $\mnt{3}$ & $\mnt{4}$ & $\mch{1}$ & $\mch{2}$ \\
  \hline \hline
  $800$ & $812$ & $925$ & $2030$ & $809$ & $2030$ \\
  \hline
\end{tabular}
\vskip 0.4cm
\begin{tabular}{|c|c|c|c|c|}
  \hline
  $m_{h^0}$ & $m_{H^0}$ & $m_{A^0}$ & $m_{H^+}$ \\
  \hline \hline
  $124.9$  & $4970$ & $4970$ & $4997$ \\
  \hline
\end{tabular}
\vskip 0.4cm
\begin{tabular}{|c|c|c|c|c|c|c|}
  \hline
  $\msg$ & $\msu{1}$ & $\msu{2}$ & $\msu{3}$ & $\msu{4}$ & $\msu{5}$ & $\msu{6}$ \\
  \hline \hline
  $2934$ & $1231$ & $2986$ & $3431$ & $3656$ & $4491$ & $4493$ \\
  \hline
\end{tabular}
\vskip 0.4cm
\begin{tabular}{|c|c|c|c|c|c|}
  \hline
 $\msd{1}$ & $\msd{2}$ & $\msd{3}$ & $\msd{4}$ & $\msd{5}$ & $\msd{6}$ \\
  \hline \hline
  $836$ & $3272$ & $3416$ & $3654$ & $4489$ & $4492$ \\
  \hline
\end{tabular}
\vskip 0.4cm
\begin{tabular}{|c|c|c|c|c|c|c|c|c|}
  \hline
  $m_{\sneut_1}$ &  $m_{\sneut_2}$ &  $m_{\sneut_3}$ &  $m_{\ti l_1}$ &  $m_{\ti l_2}$ 
                        &  $m_{\ti l_3}$ &  $m_{\ti l_4}$ &  $m_{\ti l_5}$ &  $m_{\ti l_6}$ \\
  \hline \hline
   $1506$ &  $1507$ &  $1582$ &  $1495$ &  $1496$ &  $1509$ &  $1509$ &  $1564$ &  $1652$ \\
  \hline
\end{tabular}
\end{center}
\label{physmasses}
\end{table}
%

%
\begin{table}[h!]
\caption{Flavour decompositions of the mass eigenstates $\su_{1,2,3}$ and $\sd_{1,2,3}$ for the scenario 
         of Table~\ref{table2}. Shown are the expansion coefficients of the mass eigenstates in terms 
         of the flavour eigenstates. Imaginary parts of the coefficients are negligibly small.}
\begin{center}
\begin{tabular}{|c|c|c|c|c|c|c|c|}
  \hline
  & $\su_L$ & $\sca_L$ & $\sto_L$ & $\su_R$ & $\sca_R$ & $\sto_R$ \\
  \hline
  $\su_1$  & $0$ &  $0.0016$ & $0.0992$ & $0$ & $-0.4090$ & $-0.9071$ \\
  \hline 
  $\su_2$  &  $-0.0012$ &  $-0.0070$ & $-0.0225$ & $0$ & $0.9104$ & $-0.4130$ \\
  \hline 
  $\su_3$  & $0.0660$ & $0.2921$ & $0.9491$ & $0$ & $0.0607$ & $0.0770$ \\
  \hline  \hline
  & $\sd_L$ & $\ss_L$ & $\sbo_L$ & $\sd_R$ & $\ss_R$ & $\sbo_R$ \\
  \hline
  $\sd_1$  & $0$ & $0$ &  $0.0059$ & $0$ & $0.4057$ & $0.9140$ \\
  \hline 
  $\sd_2$  & $0$ &  $0.0059$ & $0.0289$ & $0$ & $-0.9137$ & $0.4054$ \\
  \hline 
  $\sd_3$  & $0$ & $0.2898$ & $0.9566$ & $0$ & $0.0245$ & $-0.0172$ \\
  \hline
\end{tabular}
\end{center}
\label{flavourdecomp}
\end{table}
%

%

\begin{figure*}[h!]
\centering
 \subfigure[]{
   { \mbox{\hspace*{-1.0cm} \resizebox{7.5cm}{!}{\includegraphics{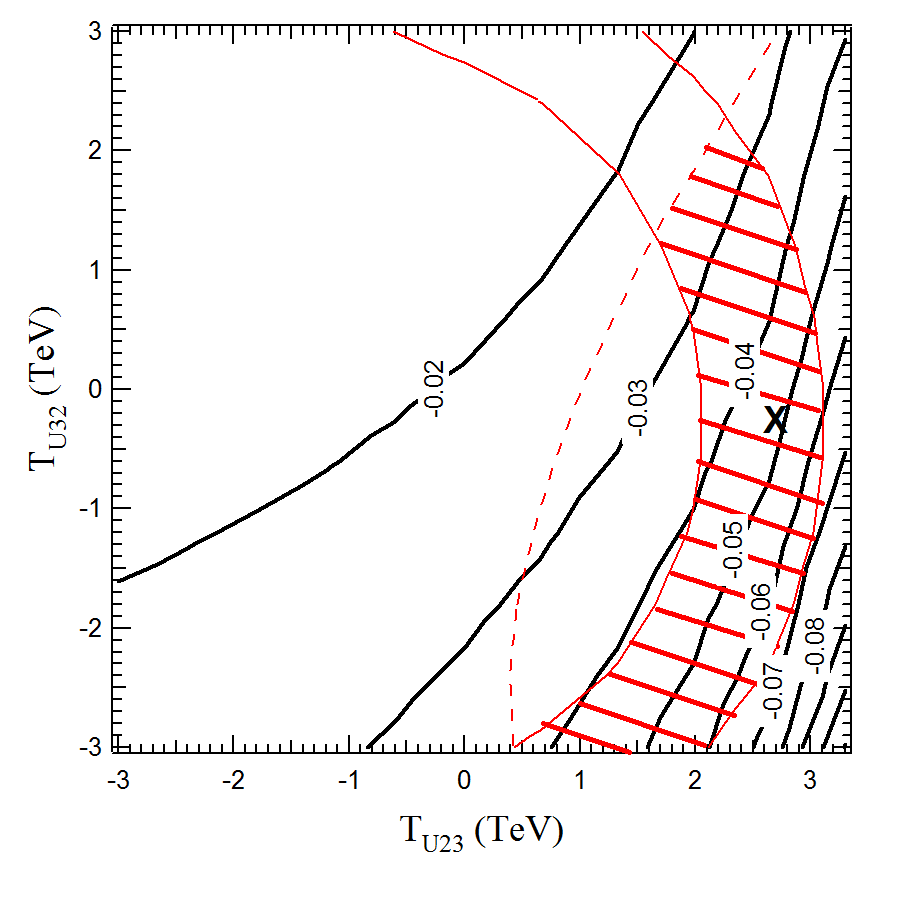}} \hspace*{0cm}}}
   \label{C'7_TU23TU32}} 
 \subfigure[]{
   { \mbox{\hspace*{0cm} \resizebox{7.5cm}{!}{\includegraphics{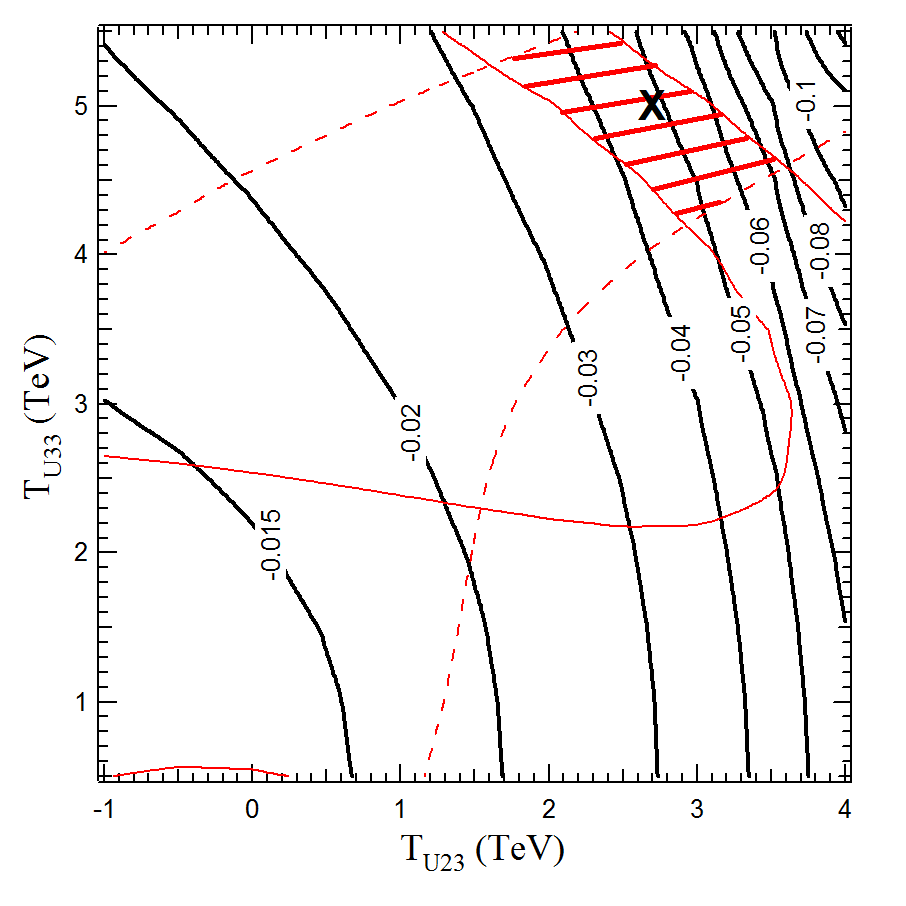}} \hspace*{0cm}}}
   \label{C'7_TU23TU33}}\\
 \subfigure[]{
   { \mbox{\hspace*{-1.0cm} \resizebox{7.5cm}{!}{\includegraphics{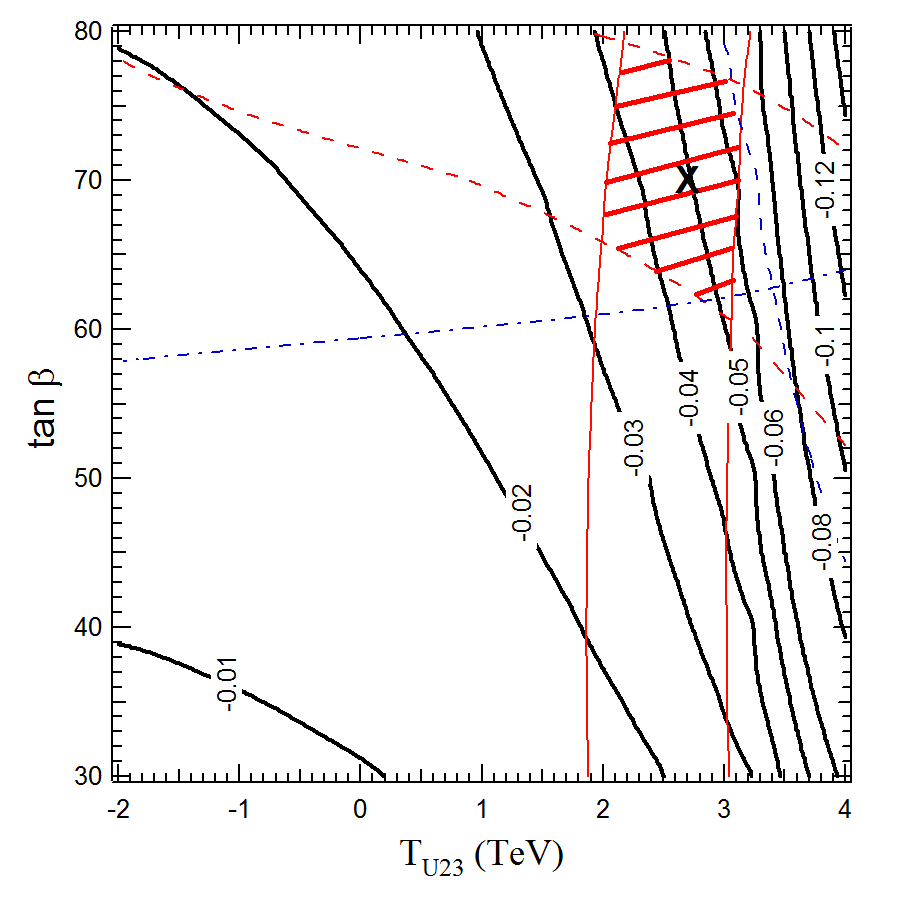}} \hspace*{0cm}}}
  \label{C'7_TU23tanb}}
 \subfigure[]{
   { \mbox{\hspace*{0cm} \resizebox{7.5cm}{!}{\includegraphics{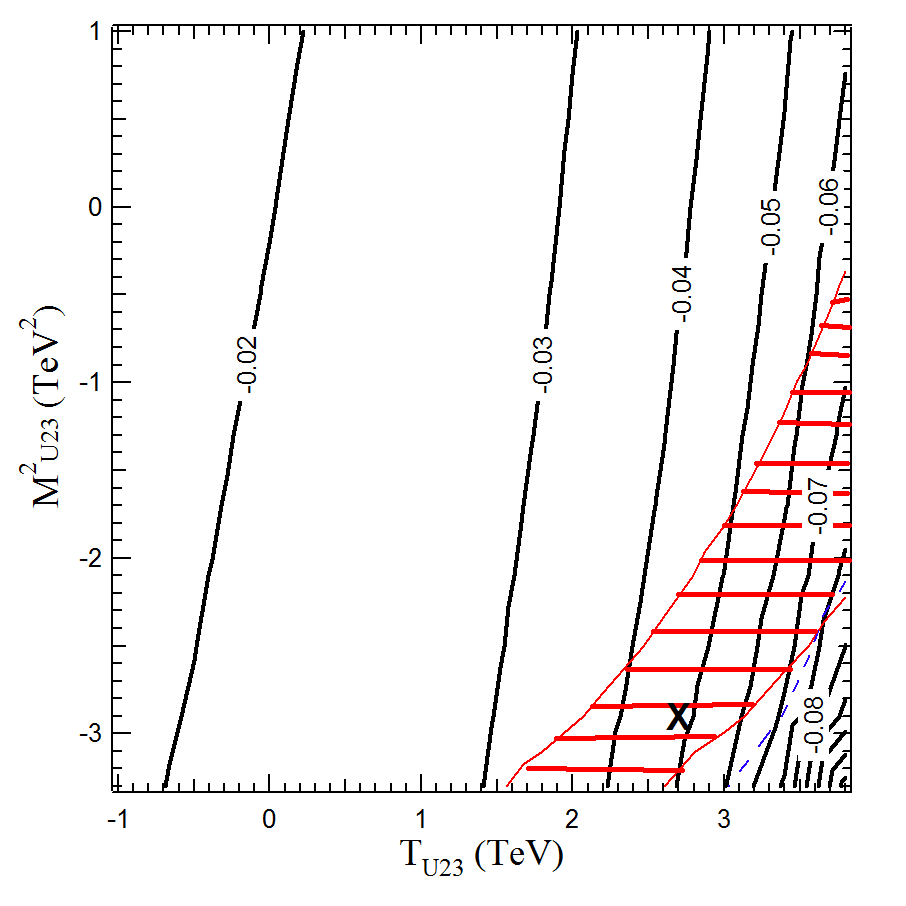}} \hspace*{0cm}}}
  \label{C'7_TU23M2U23}}
\caption{
     Contour plots of $Re(C'_7(\mu_b))$ around the benchmark point P1 in the parameter planes of
     (a) $T_{U 23}$ - $T_{U 32}$, (b) $T_{U 23}$ - $T_{U 33}$, (c) $T_{U 23}$ - $tan\beta$, and
     (d) $T_{U 23}$ - $M^2_{U 23}$. 
     The parameters other than the shown ones in each plane are fixed as in Table~\ref{table2}.
     The "X" marks P1 in the plots. 
     The red hatched region satisfies all the constraints in Appendix A. 
     The red solid lines, the blue dashed lines, the red dashed lines and the blue dash-dotted lines 
     show the $m_{h^0}$ bound, the ${\rm B}(b \to s \gamma)$ bound, the ${\rm B}(B_s \to \mu^+ \mu^-)$ bound, 
     and the $m_{\sd_1}$ bound, respectively.
%
%
     }
\label{fig_C'7_1}
\end{figure*}

\begin{figure*}[h!]
\centering
 \subfigure[]{
   { \mbox{\hspace*{-1cm} \resizebox{7.5cm}{!}{\includegraphics{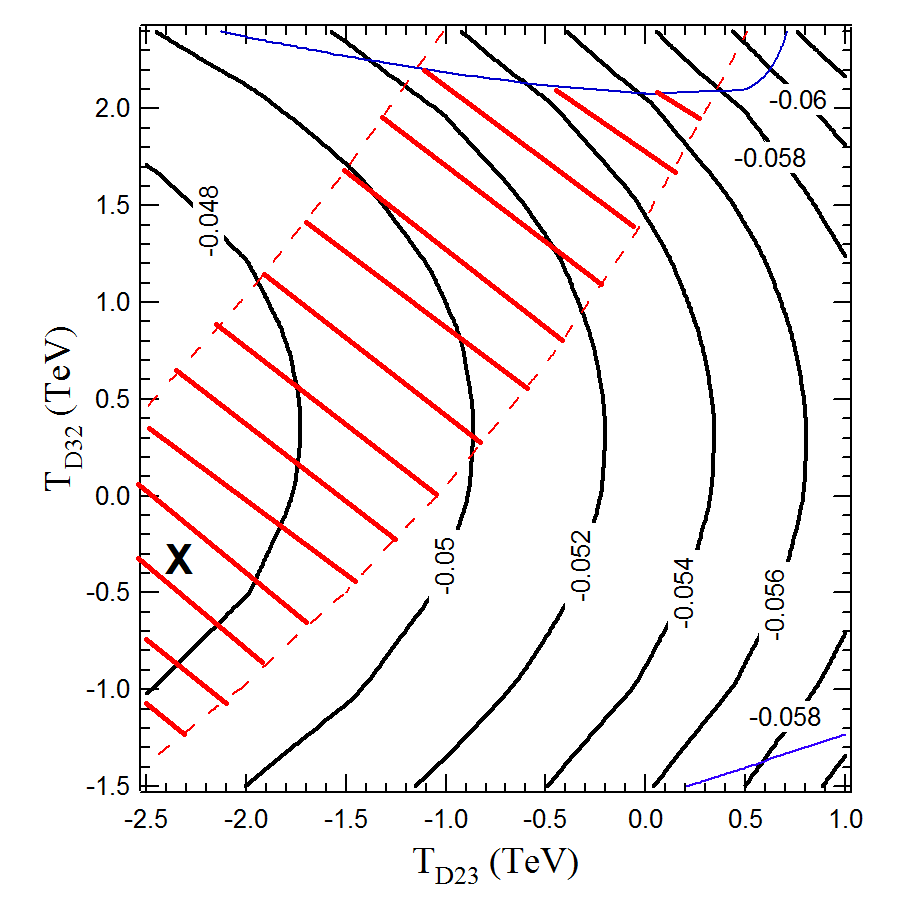}} \hspace*{0cm}}}
  \label{C'7_TD23TD32}}
 \subfigure[]{
   { \mbox{\hspace*{0cm} \resizebox{7.5cm}{!}{\includegraphics{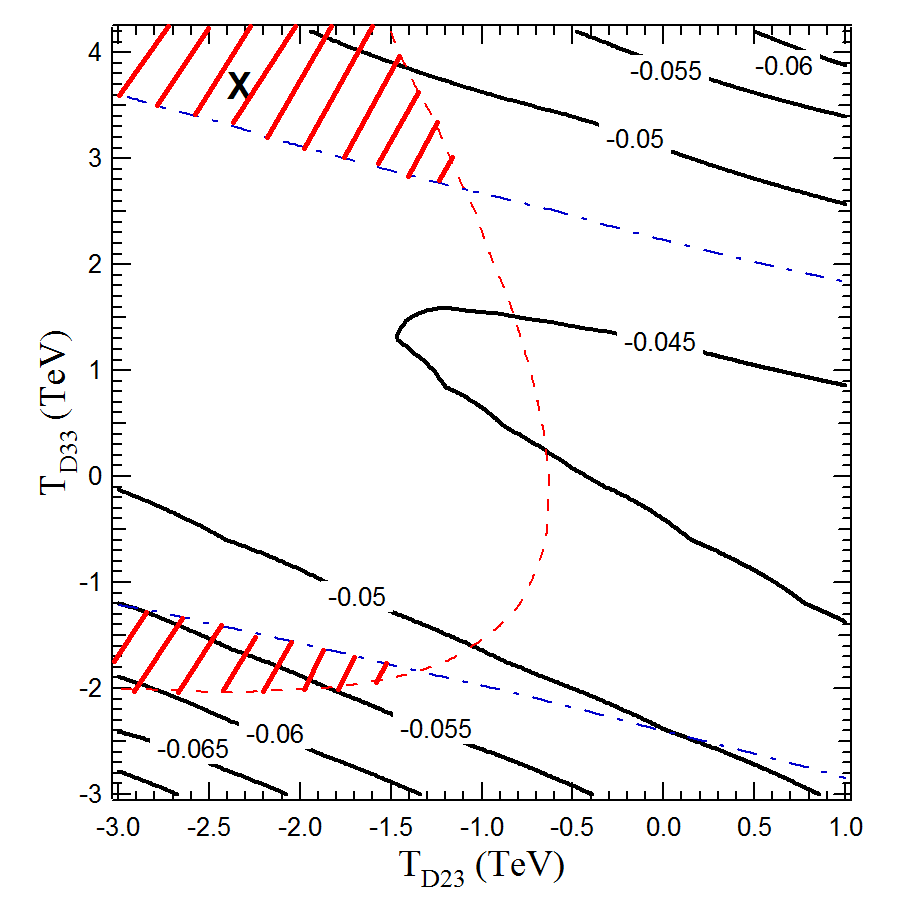}} \hspace*{0cm}}}
  \label{C'7_TD23TD33}}\\
 \subfigure[]{
   { \mbox{\hspace*{-1cm} \resizebox{7.5cm}{!}{\includegraphics{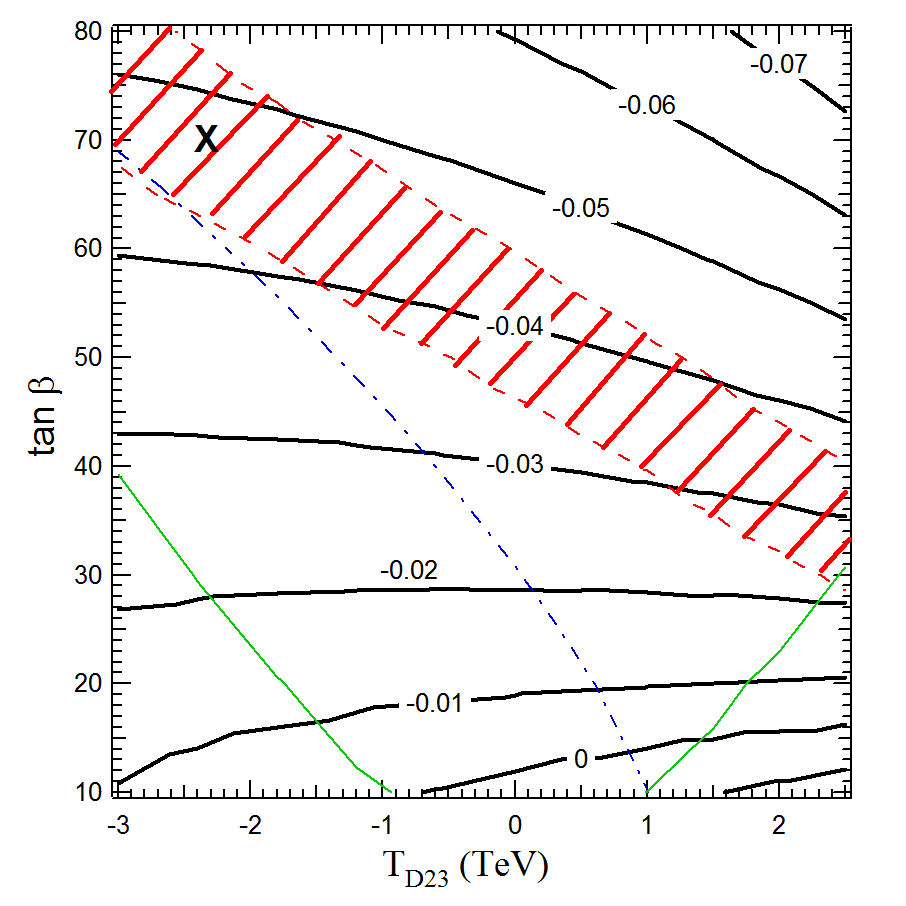}} \hspace*{0cm}}}
  \label{C'7_TD23tanb}}
 \subfigure[]{
   { \mbox{\hspace*{0cm} \resizebox{7.5cm}{!}{\includegraphics{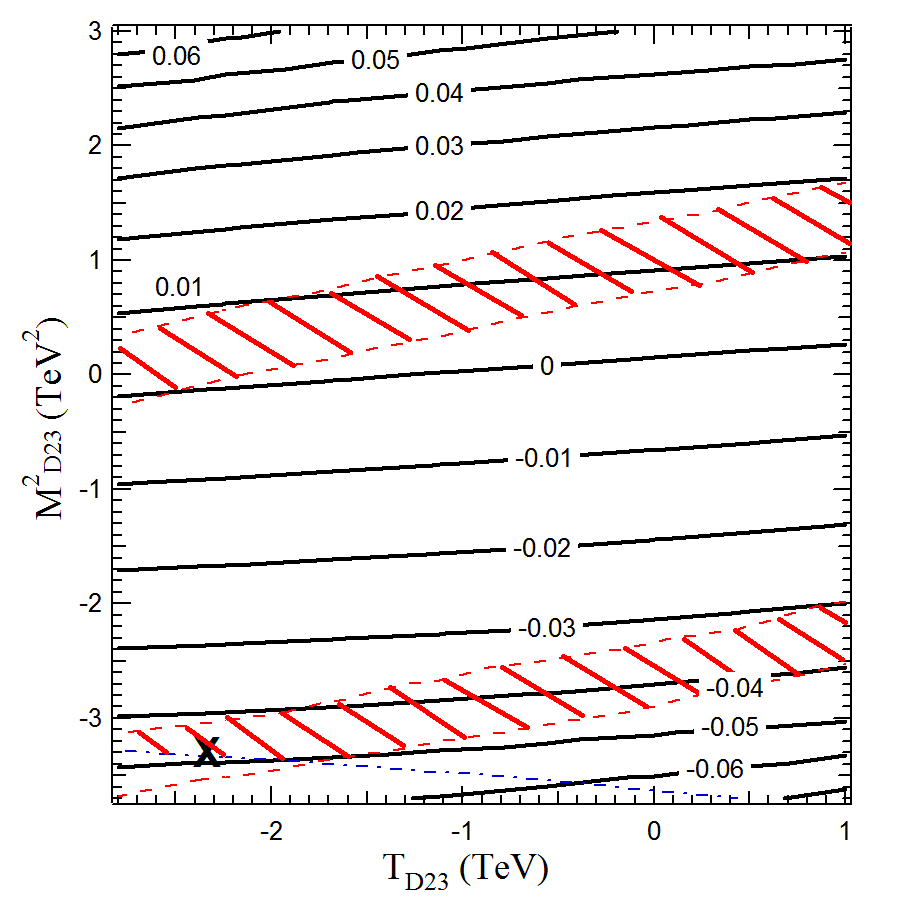}} \hspace*{0cm}}}
  \label{C'7_TD23M2D23}}
\caption{
     Contour plots of $Re(C'_7(\mu_b))$ around the benchmark point P1 in the parameter planes of
     (a) $T_{D 23}$ - $T_{D 32}$, (b) $T_{D 23}$ - $T_{D 33}$, (c) $T_{D 23}$ - $tan\beta$, and 
     (d) $T_{D 23}$ - $M^2_{D 23}$. 
     The parameters other than the shown ones in each plane are fixed as in Table~\ref{table2}.
     The "X" marks P1 in the plots. 
     The red hatched region satisfies all the constraints in Appendix A. 
     The definitions of the bound lines are the same as in Fig.~\ref{fig_C'7_1}. 
     In addition to these the blue solid lines and the green solid lines show 
     the $\Delta M_{B_s}$ bound and the vacuum stability bound on $T_{D23}$, respectively.
     }
\label{fig_C'7_2}
\end{figure*}

In Figs. \ref{fig_C'7_1} and \ref{fig_C'7_2} we show contours of $Re(C'_7(\mu_b))$ around 
the benchmark point P1 in various parameter planes. 
Fig.~\ref{C'7_TU23TU32} shows contours of $Re(C'_7(\mu_b))$ in the $T_{U23}$-$T_{U32}$ plane. 
We see that $Re(C'_7(\mu_b))$ is sensitive to both $T_{U23}$ and $T_{U32}$, especially to $T_{U23}$, 
increases quickly with the increase of $T_{U23}$ and $T_{U32} (< 0)$, as is expected, and can be as 
large as about -0.07 in the allowed region. We also see that it is large 
($-0.07 \lsim Re(C'_7(\mu_b)) \lsim -0.04$) respecting all the constraints in a significant 
part of this parameter plane. 
%
From Fig.~\ref{C'7_TU23TU33} we see that $Re(C'_7(\mu_b))$ is also fairly sensitive to $T_{U33}$
and can be as large as $\sim -0.08$. 
%
From Fig.~\ref{C'7_TU23tanb} we find that $Re(C'_7(\mu_b))$ is very sensitive to $tan\b$, 
especially for large $T_{U23} > 0$, as expected, and can be as large as $\sim -0.07$. 
%
As can be seen in Fig.~\ref{C'7_TU23M2U23}, $Re(C'_7(\mu_b))$ is sensitive to $M^2_{U 23}$, 
especially for large $T_{U23} \gsim 2.5$ TeV, as expected, and is large 
($-0.08 \lsim Re(C'_7(\mu_b)) \lsim -0.04$) respecting all the constraints in a 
significant part of this parameter plane.
%
%
Fig.~\ref{C'7_TD23TD32} shows contours of $Re(C'_7(\mu_b))$ in the $T_{D23}$-$T_{D32}$ plane. 
It is fairly sensitive to $T_{D23}$ and mildly dependent on $T_{D32}$ as is expected 
partly from the contribution of Fig.~\ref{bL2sRgamg_sg_loop} (see Eq.(\ref{C7})), can be as large as 
$\sim -0.06$ in the allowed region, and is large ($-0.058 \lsim Re(C'_7(\mu_b)) \lsim -0.046$) 
respecting all the constraints in a significant part of this parameter plane.
%
From Fig.~\ref{C'7_TD23TD33} we see that $Re(C'_7(\mu_b))$ is also rather sensitive to $T_{D33}$ and 
can be as large as $\sim -0.06$ in the allowed region.
%
As can be seen in Fig.~\ref{C'7_TD23tanb}, $Re(C'_7(\mu_b))$ is very sensitive to $tan\b$ 
and also sensitive to $T_{D23}$ for large $tan\b \gsim 70$, as expected, and is 
sizable ($-0.05 \lsim Re(C'_7(\mu_b)) \lsim -0.04$) respecting all the constraints in a 
significant part of this parameter plane.
%
From Fig.~\ref{C'7_TD23M2D23} we find that $Re(C'_7(\mu_b))$ is very sensitive to $M^2_{D 23}$, 
and is sizable ($-0.05 \lsim Re(C'_7(\mu_b)) \lsim -0.04$) respecting all the constraints in a 
significant part of this parameter plane.

\begin{figure*}[h!]
\centering
 \subfigure[]{
   { \mbox{\hspace*{-1cm} \resizebox{7.5cm}{!}{\includegraphics{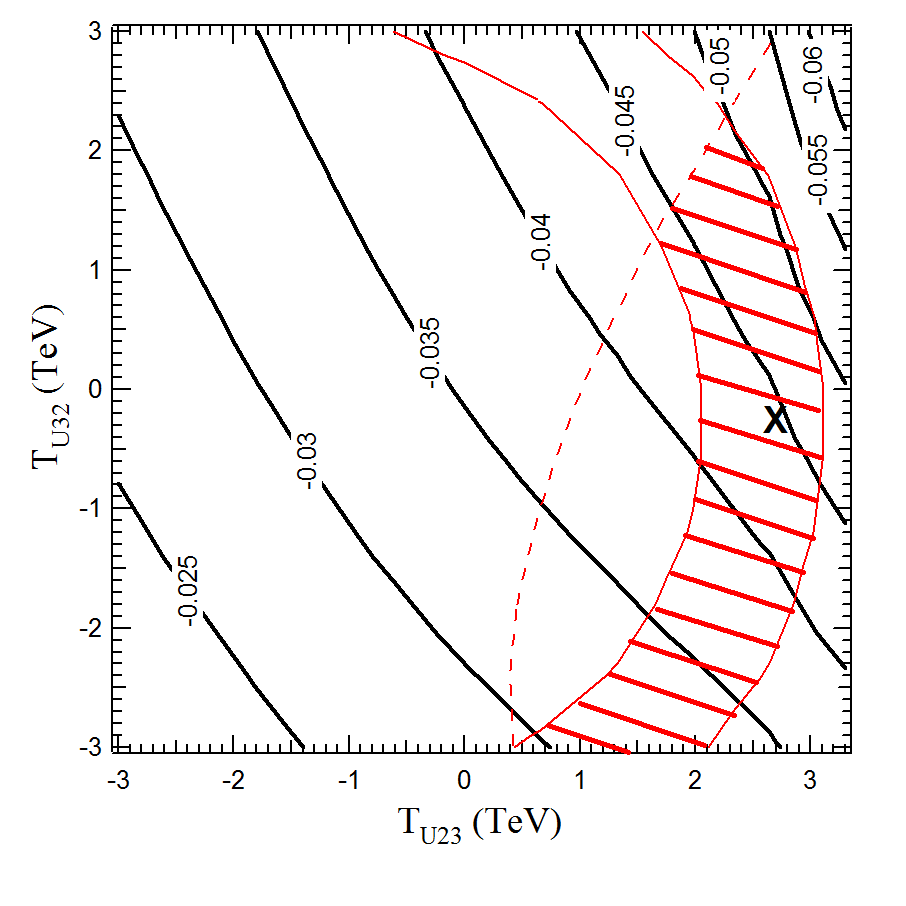}} \hspace*{0cm}}}
   \label{C7MSSM_TU23TU32}} 
 \subfigure[]{
   { \mbox{\hspace*{0cm} \resizebox{7.5cm}{!}{\includegraphics{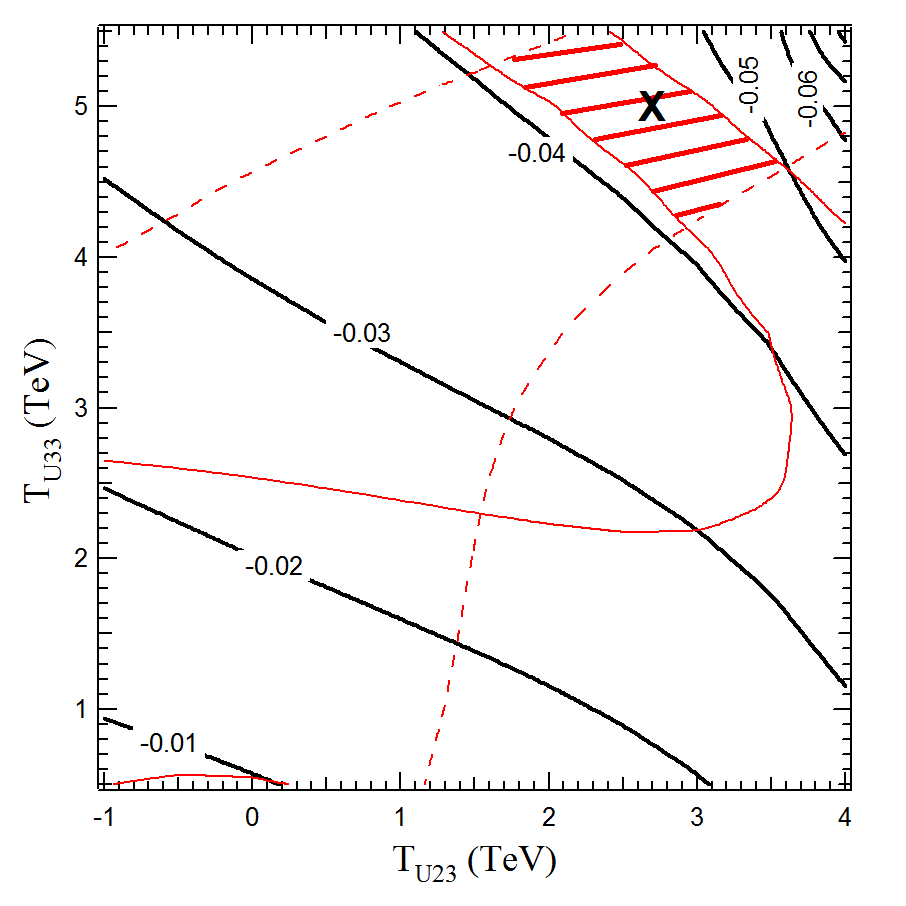}} \hspace*{0cm}}}
   \label{C7MSSM_TU23TU33}}\\
 \subfigure[]{
   { \mbox{\hspace*{-1cm} \resizebox{7.5cm}{!}{\includegraphics{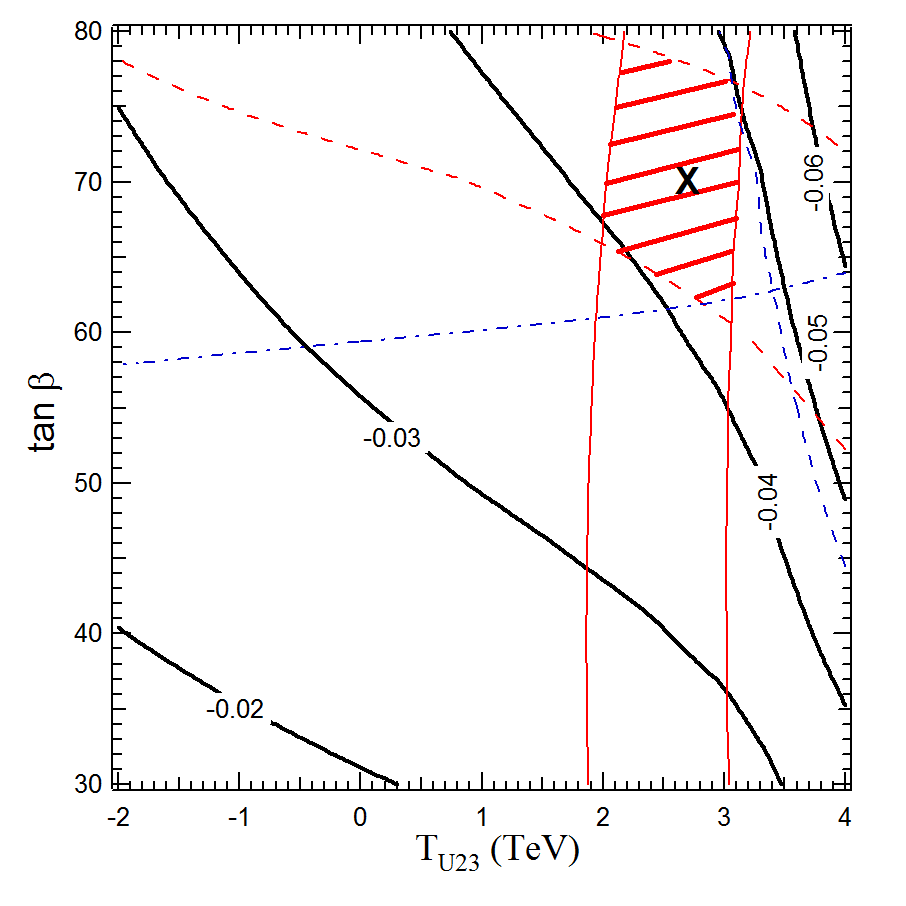}} \hspace*{0cm}}}
  \label{C7MSSM_TU23tanb}}
 \subfigure[]{
   { \mbox{\hspace*{0cm} \resizebox{7.5cm}{!}{\includegraphics{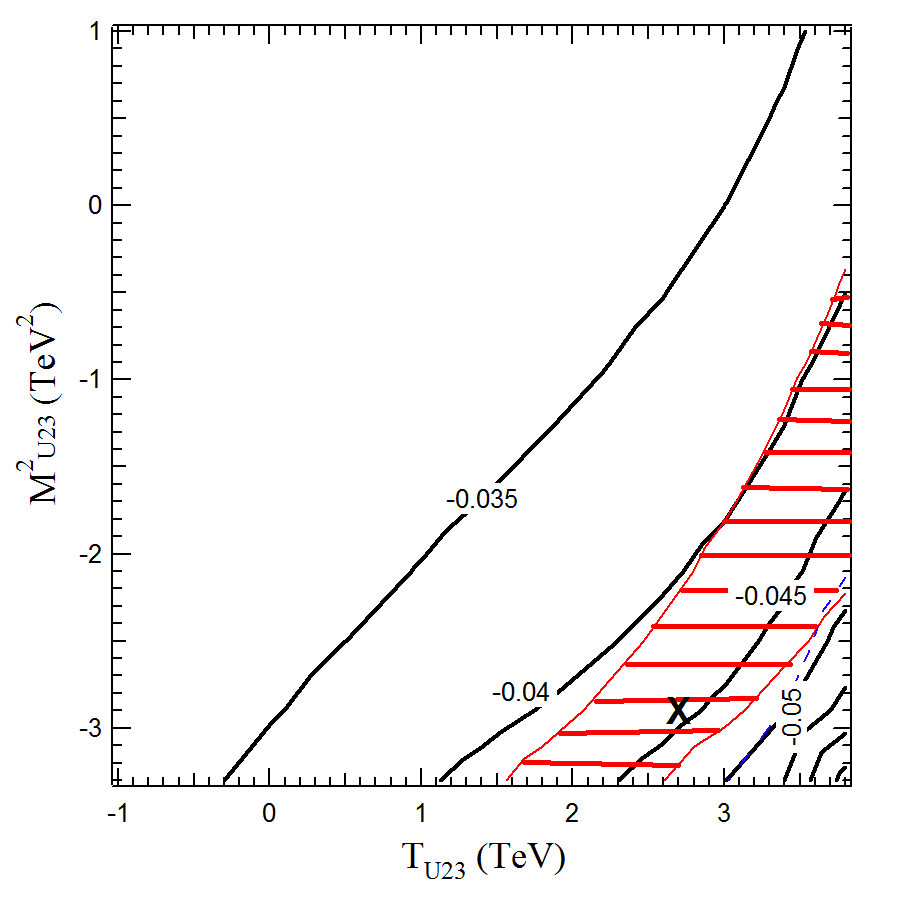}} \hspace*{0cm}}}
  \label{C7MSSM_TU23M2U23}}
\caption{
     Contour plots of $Re(C^{\rm MSSM}_7(\mu_b))$ around the benchmark point P1 in the parameter planes of
     (a) $T_{U 23}$ - $T_{U 32}$, (b) $T_{U 23}$ - $T_{U 33}$, (c) $T_{U 23}$ - $tan\beta$, and
     (d) $T_{U 23}$ - $M^2_{U 23}$. 
     The parameters other than the shown ones in each plane are fixed as in Table~\ref{table2}.
     The "X" marks P1 in the plots. 
     The red hatched region satisfies all the constraints in Appendix A. 
     The definitions of the bound lines are the same as those in Fig.~\ref{fig_C'7_1}. 
     }
\label{fig_C7MSSM_1}
\end{figure*}

\begin{figure*}[h!]
\centering
 \subfigure[]{
   { \mbox{\hspace*{-1cm} \resizebox{7.5cm}{!}{\includegraphics{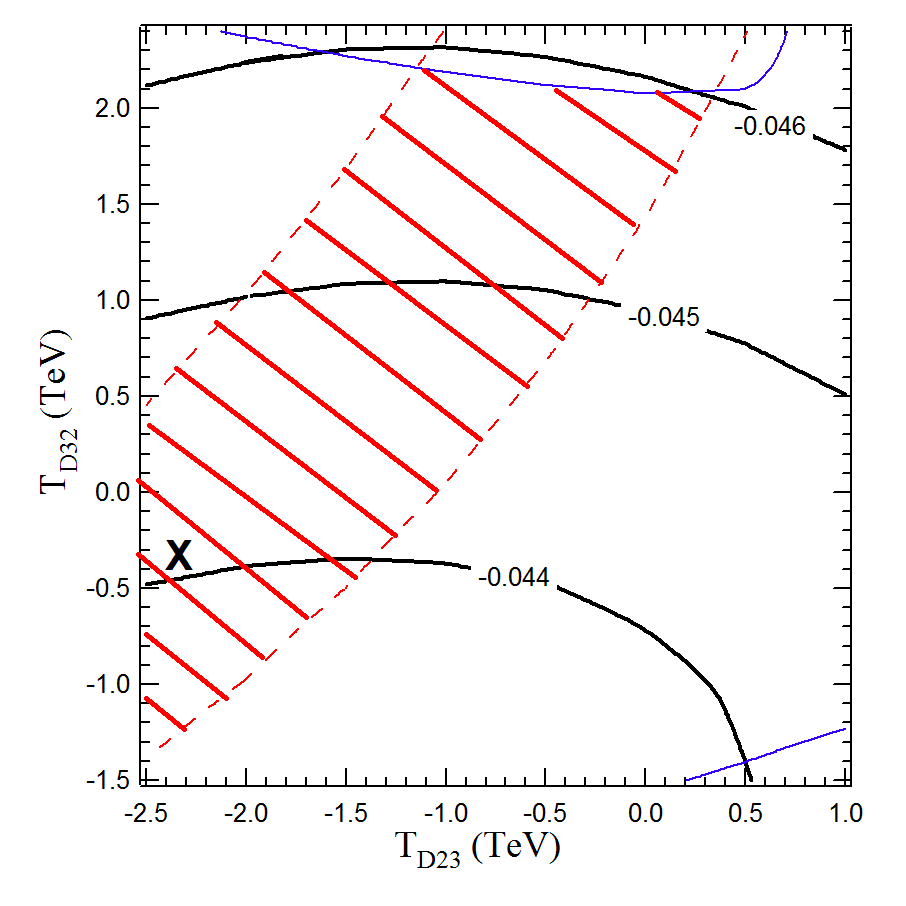}} \hspace*{0cm}}}
  \label{C7MSSM_TD23TD32}}
 \subfigure[]{
   { \mbox{\hspace*{0cm} \resizebox{7.5cm}{!}{\includegraphics{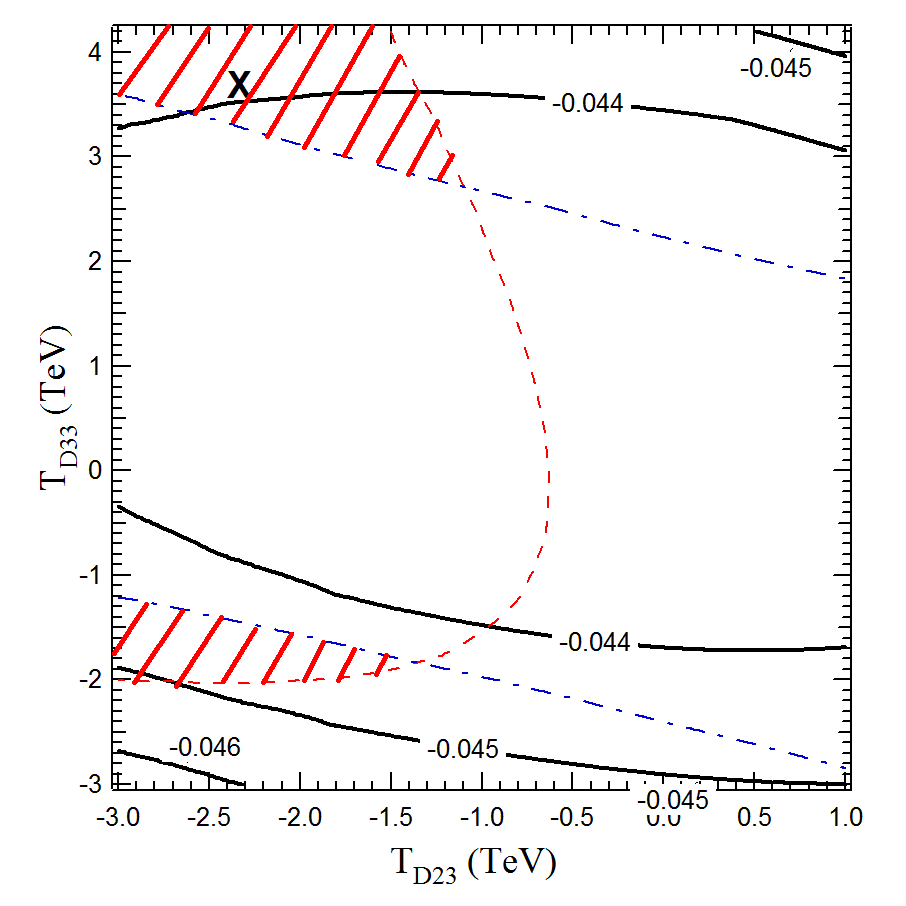}} \hspace*{0cm}}}
  \label{C7MSSM_TD23TD33}}\\
 \subfigure[]{
   { \mbox{\hspace*{-1cm} \resizebox{7.5cm}{!}{\includegraphics{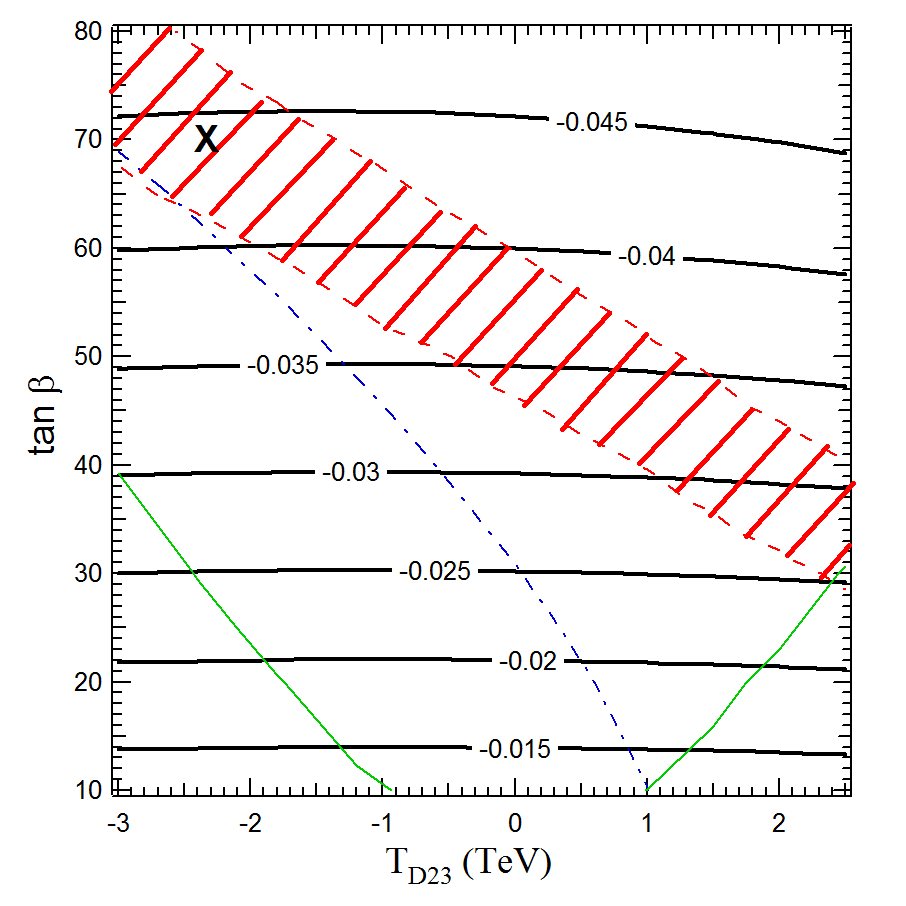}} \hspace*{0cm}}}
  \label{C7MSSM_TD23tanb}}
 \subfigure[]{
   { \mbox{\hspace*{0cm} \resizebox{7.5cm}{!}{\includegraphics{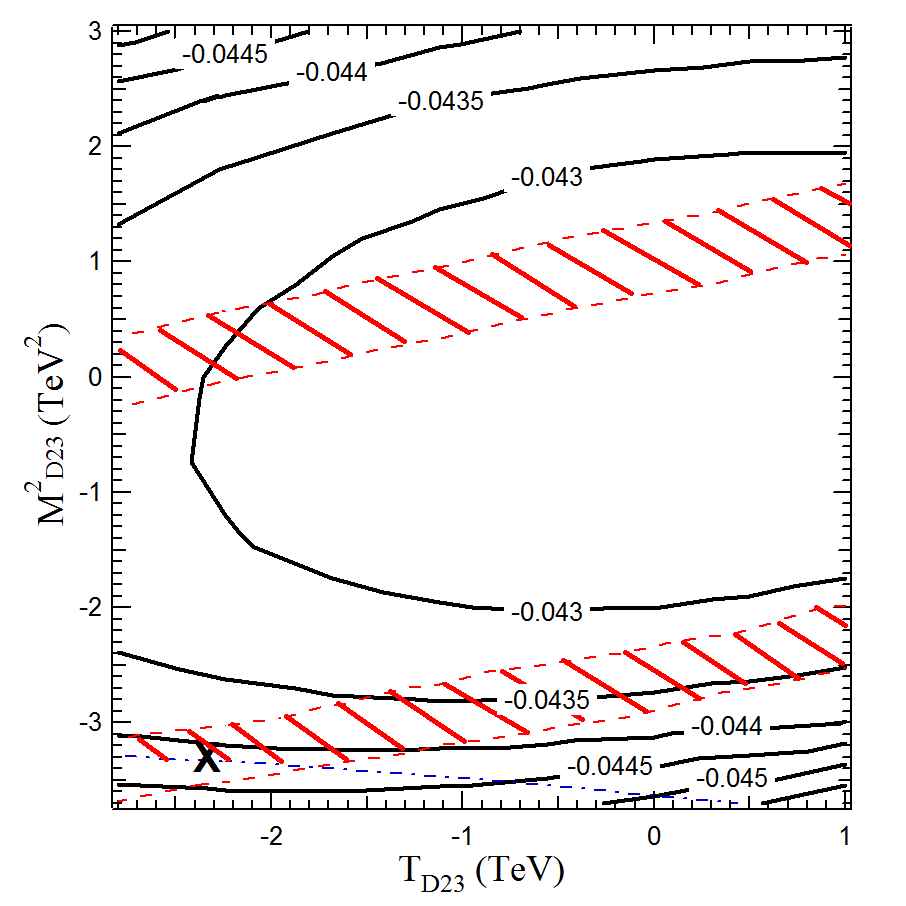}} \hspace*{0cm}}}
  \label{C7MSSM_TD23M2D23}}
\caption{
     Contour plots of $Re(C^{\rm MSSM}_7(\mu_b))$ around the benchmark point P1 in the parameter planes of
     (a) $T_{D 23}$ - $T_{D 32}$, (b) $T_{D 23}$ - $T_{D 33}$, (c) $T_{D 23}$ - $tan\beta$, and 
     (d) $T_{D 23}$ - $M^2_{D 23}$. 
     The parameters other than the shown ones in each plane are fixed as in Table~\ref{table2}.
     The "X" marks P1 in the plots. 
     The red hatched region satisfies all the constraints in Appendix A. 
     The definitions of the bound lines are the same as those in Fig.~\ref{fig_C'7_2}.
     }
\label{fig_C7MSSM_2}
\end{figure*}

In Figs. \ref{fig_C7MSSM_1} and \ref{fig_C7MSSM_2} we show contour plots of $Re(C^{\rm MSSM}_7(\mu_b))$ 
(i.e. the MSSM contributions to $Re(C_7(\mu_b))$) around the benchmark point P1 in various parameter 
planes. 
\indent Fig.~\ref{C7MSSM_TU23TU32} shows contours of $Re(C^{\rm MSSM}_7(\mu_b))$ in the $T_{U23}$-$T_{U32}$ plane.
We see that $Re(C^{\rm MSSM}_7(\mu_b))$ is sensitive to $T_{U23}$ and $T_{U32}$: 
$|Re(C^{\rm MSSM}_7(\mu_b))|$ quickly increases with the increase of $T_{U23}$ and $T_{U32}$ as is expected. 
We find also that $Re(C^{\rm MSSM}_7(\mu_b))$ can be as large as about -0.05 in the allowed region 
and is sizable ($-0.05 \lsim Re(C^{\rm MSSM}_7(\mu_b)) \lsim -0.04$) respecting all the constraints 
in a significant part of this parameter plane.
From Fig.~\ref{C7MSSM_TU23TU33} we see that $Re(C^{\rm MSSM}_7(\mu_b))$ is very sensitive also 
to $T_{U33}$ (see Fig.~\ref{TU33_C7MSSM} also), quickly increases with increase of $T_{U33}$ 
as is expected partly from the important contribution of Fig.~\ref{bR2sLgamg_chargino_loop_b} 
(see Eq.(\ref{C7})), and can be as large as about -0.05 in the allowed region. It is sizable 
($-0.05 \lsim Re(C^{\rm MSSM}_7(\mu_b)) \lsim -0.04$) respecting all the constraints in a significant 
part of this parameter plane.
%
From Fig.~\ref{C7MSSM_TU23tanb} we find that $Re(C^{\rm MSSM}_7(\mu_b))$ is very sensitive 
to $tan\b$ and $T_{U23}$ as expected, quickly increases with increase 
of $tan\b$ and $T_{U23}(>0)$, and can be as large as $\sim -0.05$ in the allowed region. 
%
As can be seen in Fig.~\ref{C7MSSM_TU23M2U23}, $Re(C^{\rm MSSM}_7(\mu_b))$ is sensitive to 
$M^2_{U 23}$ and $T_{U23}$ increasing with the increase of $M^2_{U 23} (< 0)$ and 
$T_{U23} (> 0)$ as expected, and is large ($-0.05 \lsim Re(C^{\rm MSSM}_7(\mu_b)) \lsim -0.04$) 
respecting all the constraints in a significant part of this parameter plane.
%

%
Fig.~\ref{C7MSSM_TD23TD32} shows contours of $Re(C^{\rm MSSM}_7(\mu_b))$ in the 
$T_{D23}$-$T_{D32}$ plane.
We see that $Re(C^{\rm MSSM}_7(\mu_b))$ is mildly dependent on $T_{D23}$ and fairly sensitive 
to $T_{D32}$ around P1 as is expected partly from the contribution of Fig.~\ref{bR2sLgamg_sg_loop} 
(see Eq.(\ref{C7})).
It can be as large as about -0.046 in the allowed region, and is sizable 
($-0.046 \lsim Re(C^{\rm MSSM}_7(\mu_b)) \,\lsim -0.044$) respecting all the constraints 
in a significant part of this parameter plane.
%
From Fig.~\ref{C7MSSM_TD23TD33} we see that $Re(C^{\rm MSSM}_7(\mu_b))$ is also fairly 
sensitive to $T_{D33}$ around P1, can be as large as about -0.045 in the allowed region, 
and is sizable ($-0.045 \lsim Re(C^{\rm MSSM}_7(\mu_b)) \lsim -0.044$) respecting all the 
constraints in a significant part of this parameter plane.
%
From Fig.~\ref{C7MSSM_TD23tanb} we find that $Re(C^{\rm MSSM}_7(\mu_b))$ is very sensitive 
to $tan\b$ quickly increasing with the increase of 
$tan\b$ as expected, can be as large as $\sim -0.05$ in the allowed region, and is 
sizable ($-0.05 \lsim Re(C^{\rm MSSM}_7(\mu_b)) \lsim -0.04$) respecting all the constraints 
in a significant part of this parameter plane.
%
As can be seen in Fig.~\ref{C7MSSM_TD23M2D23}, $Re(C^{\rm MSSM}_7(\mu_b))$ is mildly dependent on 
$M^2_{D 23}$ around this benchmark point P1, can be as large as about -0.044
in the allowed region, and is sizable ($-0.044 \lsim Re(C^{\rm MSSM}_7(\mu_b)) \lsim -0.043$) 
respecting all the constraints in a significant part of this parameter plane.\\

As the gluino is very heavy ($\sim$ 3 TeV) around the reference point P1 (see Table \ref{physmasses}), 
the down-type squark - gluino loop contributions to $Re(C'_7(\mu_b))$ and $Re(C^{\rm MSSM}_7(\mu_b))$ are 
suppressed there, which partly explains the rather mild dependences of $Re(C'_7(\mu_b))$ and 
$Re(C^{\rm MSSM}_7(\mu_b))$ on the down-type squark parameters $T_{D23}$, $T_{D32}$, $T_{D33}$ around P1
as is seen in Fig.~\ref{fig_C'7_2} and Fig.~\ref{fig_C7MSSM_2}, respectively.
\\

Before closing this section we comment on the renormalization scale dependence of 
the WCs $C^{\rm MSSM}_7(\mu_b)$ and $C'_7(\mu_b)$. For the reference scenario P1 
we have the following result at LO:\\
$(Re(C_7(\mu_b/2)), Im(C_7(\mu_b/2)) = (-0.405, -4.04\cdot 10^{-4})$, 
$(Re(C^{\rm MSSM}_7(\mu_b/2)), Im(C^{\rm MSSM}_7\\
(\mu_b/2)) = (-0.0379, -4.04\cdot 10^{-4})$ and 
$(Re(C'_7(\mu_b/2)), Im(C'_7(\mu_b/2)) = (-0.0350, -1.34\cdot 10^{-3})$; 
$(Re(C_7(2\mu_b)), Im(C_7(2\mu_b)) = (-0.341, -6.19\cdot 10^{-4})$, 
$(Re(C^{\rm MSSM}_7(2\mu_b)), Im(C^{\rm MSSM}_7\\
(2\mu_b)) = (-0.0499, -6.20\cdot 10^{-4})$ and 
$(Re(C'_7(2\mu_b)), Im(C'_7(2\mu_b)) = (-0.0594, -2.28\cdot 10^{-3})$, where $\mu_b = 4.8$ GeV.
%
%
We see that the scale dependence of the WCs at the b-quark mass scale is significant at LO 
in agreement with Refs. \cite{Ali_92, Ali_93, Buras_93} 
and hence that it is important to compute the WCs at higher order (NLO/NNLO) level in order 
to reduce this scale-dependence uncertainties. 
In \cite{Hurth_2011} MSSM loop contributions to the WCs $C_{7,8}(\mu_W)$ 
and $C'_{7,8}(\mu_W)$ are calculated at NLO in the MSSM with QFV.
So far, however, there is no complete NLO computation of the WCs $C_7(\mu_b)$ and 
$C'_7(\mu_b)$ in the MSSM with QFV 
\footnote{
In principle the MSSM loop contributions to $C_7(\mu_b)$ and $C'_7(\mu_b)$ at NLO 
can be obtained from $C_i(\mu_W)$ and $C'_i(\mu_W)$ (i = 1-8) calculated at NLO in 
the MSSM by using QCD RG scale evolution from the scale $\mu_W$ down to $\mu_b$ 
at NLL (next-to-leading log) level ~\cite{Buras_93}, where $C_i(\mu_W)$ and $C'_i(\mu_W)$ 
(i = 1-6) are the Wilson coefficients of the four-quark operators.
}.

\newpage
\section{Conclusions}
\label{sec:concl}

We have studied SUSY effects on $C_7(\mu_b)$ and $C'_7(\mu_b)$ 
which are the Wilson coefficients for $b \to s \gamma$ at b-quark 
mass scale $\mu_b$ and are closely related to radiative $B$-meson decays.  
The SUSY-loop contributions to the $C_7(\mu_b)$ and $C'_7(\mu_b)$ are calculated 
at LO in the Minimal Supersymmetric Standard Model with general quark-flavour 
violation. For the first time we have performed a systematic MSSM parameter 
scan for the WCs $C_7(\mu_b)$ and $C'_7(\mu_b)$ respecting all the relevant 
constraints, i.e. the theoretical 
constraints from vacuum stability conditions and the 
experimental constraints, such as those from $K$- and $B$-meson data and 
electroweak precision data, as well as recent limits on 
SUSY particle masses and the 125 GeV Higgs boson data from LHC experiments. 
From the parameter scan, we have found the following: 
\begin{itemize}
\item  The MSSM contribution to Re($C_7(\mu_b)$) can be as large as $\sim \pm 0.05$ 
  which could correspond to about 3$\sigma$ significance of NP (New Physics) signal 
  in future Belle II and LHCb Upgrade experiments.
\item  The MSSM contribution to Re($C'_7(\mu_b)$) can be as large as $\sim -0.08$ 
  which could correspond to about 4$\sigma$ significance of NP signal 
  in future Belle II and LHCb Upgrade experiments.
\item  These large MSSM contributions to the WCs are mainly 
  due to (i) large scharm-stop mixing and large scharm/stop involved 
  trilinear couplings $T_{U23}$, $T_{U32}$ and $T_{U33}$, (ii) large 
  sstrange-sbottom mixing and large sstrange-sbottom involved 
  trilinear couplings $T_{D23}$, $T_{D32}$ and $T_{D33}$, and 
  (iii) large bottom Yukawa coupling $Y_b$ for large $\tan\beta$ 
  and large top Yukawa coupling $Y_t$.
\end{itemize}

Moreover, we have pointed out the following:
\begin{itemize}
\item  It is very important to reduce the (theoretical and experimental) errors of 
the WCs $C'_7(\mu_b)$ and $C_7^{NP}(\mu_b)$ obtained (extracted) from the future experiments at 
Belle II and the LHCb Upgrade. An improvement in precision of both theory and 
experiment by a factor about 1.5 or so would be very important in view of NP search 
(such as SUSY search). Therefore, we strongly encourage theorists and 
experimentalists to challenge this task.\\
\item  On the other hand, it is also very important to reduce the theoretical errors 
of the MSSM contributions to the WCs $C'_7(\mu_b)$ and $C_7(\mu_b)$ by performing 
higher order computations such as those at NLO/NNLO level.\\
\end{itemize}

\noindent
In case such large New Physics contributions to the WCs, i.e. such large 
deviations of the WCs from their SM values, are really observed
in the future experiments at Belle II and the LHCb Upgrade, 
this could be the imprint of QFV SUSY (the MSSM with general QFV) 
and would encourage to perform further studies of the WCs $C'_7(\mu_b)$ and 
$C_7^{\rm MSSM}(\mu_b)$ at NLO/NNLO level in this model.

%
\section*{Acknowledgments}

We would like to thank W. Porod for helpful discussions, especially for the 
permanent support concerning SPheno. 
VRVis is funded by BMVIT, BMDW, Styria, SFG and Vienna Business Agency in the 
scope of COMET - Competence Centers for Excellent Technologies (854174) which 
is managed by FFG.\\

\begin{appendix}
 
\section{Theoretical and experimental constraints}
\label{sec:constr}
The experimental and theoretical constraints taken into account in the 
present work are discussed in detail in~\cite{Eberl_17}. 
Here we list the updated constraints from $K$- and B-physics and those 
on the Higgs boson mass and couplings in Table~\ref{TabConstraints}.
For the mass of the Higgs boson $h^0$, taking the combination of the ATLAS and 
CMS measurements  $m_{h^0} = 125.09 \pm 0.24~\gev$ \cite{Higgs_mass_ATLAS_CMS} and 
adding the theoretical uncertainty of $\sim \pm 3~\gev$ ~\cite{Higgs_mass_Heinemeyer}
linearly to the experimental uncertainty at 2$\sigma$, 
we take $m_{h^0} = 125.09 \pm 3.48 ~\gev$.
The $h^0$ couplings that receive SUSY QFV effects significantly are 
$C(hbb)$ ~\cite{Eberl:h2bb}, $C(hcc)$ ~\cite{Bartl:2014bka}, 
$C(hgg)$ and $C(h\gamma\gamma)$ ~\cite{h2gagagg}
\footnote{
Precisely speaking, in principle, $C(htt)$ coupling could also 
receive SUSY QFV effects significantly. 
However, predicting the (effective) coupling $C(htt)$ at loop levels in the MSSM 
is very difficult since its theoretical definition in the context of $t\bar{t}h$ 
production at LHC is unclear ~\cite{tth@LHC}.
}. 
The measurement of $C(hcc)$ is very difficult 
due to huge QCD backgrounds at LHC; there is no significant experimental data 
on $C(hcc)$ at this moment. Hence, the relevant $h^0$ couplings to be compared 
with the LHC observations are $C(hbb)$, $C(hgg)$ and $C(h\gamma\gamma)$. 
Therefore, we list the LHC data on $C(hbb)$ ($\kappa_b$), $C(hgg)$ ($\kappa_g$) 
and $C(h\gamma\gamma)$ ($\kappa_\gamma$) in Table~\ref{TabConstraints}.\\


%
%
%

As the constraints from the decays $B\to D^{(*)}\,\tau\,\nu$ are unclear due to 
large theoretical uncertainties \cite{Bartl:2014bka}
\footnote{
As pointed out in \cite{Nierste:2008qe}, the theoretical 
predictions (in the SM and MSSM) on B$(B \to D\, l\, \nu)$ and B$(B \to D^*\, l\, \nu)$~$
(l = \tau, \mu, e)$ have potentially large theoretical uncertainties due to the 
theoretical assumptions on the form factors at the $B\,D\,W^+$ and $B\,D^*\,W^+$ 
vertices (also at the $B\,D\,H^+$  and $B\,D^*\,H^+$ vertices in the MSSM). Hence 
the constraints from these decays are unclear. 
}, 
we don't take these constraints into account in our paper.
As the issues of possible anomalies of $R(D^{(*)}) = B(B\to D^{(*)}\,\tau\,\nu)/B(B\to D^{(*)}\,\ell\,\nu)$ 
with $\ell = e \ \mbox{or} \ \mu$ and $R_{K^{(*)}} = B(B\to K^{(*)}\,e^+\, e^-)/B(B\to K^{(*)}\,\mu^+\,\mu^-)$
are not yet settled \cite{B@ICHEP2020, HFAG2019}, we don't take the constraints from these ratios into 
account either.
In \cite{Dedes} the QFV decays $t \to q h^0$ with $q = u, c$, have 
been studied in the general MSSM with QFV. It is found that these decays cannot 
be visible at the current and high luminosity LHC runs due to the very small 
decay branching ratios B($t \to q h^0$), giving no significant constraint on the 
$\tilde c - \tilde t$ mixing.\\ 
We comment on the very recent data on the anomalous magnetic moment 
of muon $a_\mu$ from the Fermilab experiment \cite{a_muon_Fermilab}. 
The Fermilab data has been combined with the previous BNL data \cite{a_muon_BNL} resulting 
in 4.2$\sigma$ discrepancy between the experimental data and the SM prediction
\footnote{
It is worth noting that according to the recent computation of the leading 
order hadronic vacuum polarization contribution to $a_\mu$ using lattice 
QCD \cite{BMW_collab}, the discrepancy between the experimental data and the SM 
prediction is only about 1.6 $\sigma$.
}.
In our scenario with heavy sleptons/sneutrinos with masses of about 1.5 TeV  
the MSSM loop contributions to $a_\mu$ are so small that they can not 
explain the discrepancy between the new data and the SM prediction.
Therefore, in the context of our scenario, this discrepancy should be 
explained by the loop contributions of another new physics coexisting 
with SUSY.
\\

In addition to these we also require our scenarios to be 
consistent with the following experimental constraints:

\begin{table*}[h!]
\footnotesize{
\caption{
Constraints on the MSSM parameters from the $K$- and $B$-meson data 
relevant mainly for the mixing between the second and the third generations of 
squarks and from the data on the $h^0$ mass and couplings $\kappa_b$, $\kappa_g$, 
$\kappa_\gamma$. The fourth column shows constraints at $95 \%$ CL obtained by 
combining the experimental error quadratically with the theoretical uncertainty, 
except for $B(K^0_L \to \pi^0 \nu \bar{\nu})$, $m_{h^0}$ and $\kappa_{b,g,\gamma}$.
}
\begin{center}
\begin{tabular}{|c|c|c|c|}
    \hline
    Observable & Exp.\ data & Theor.\ uncertainty & \ Constr.\ (95$\%$CL) \\
    \hline\hline
    &&&\\
    $10^3\times|\epsilon_K|$ & $2.228 \pm 0.011$ (68$\%$ CL)~\cite{PDG2020} 
    & $\pm 0.28$ (68$\%$ CL)~\cite{epsK_DMK_SM} &
    $2.228 \pm 0.549$\\
    $10^{15}\times\Delta M_K$ [GeV] & $3.484\pm 0.006$ (68$\%$ CL)~\cite{PDG2020} 
    & $\pm 1.2 $ (68$\%$ CL)~\cite{epsK_DMK_SM} &
    $3.484 \pm 2.352$\\
    $10^{9}\times$B($K^0_L \to \pi^0 \nu \bar{\nu}$) & $< 3.0$ (90$\%$ CL)~\cite{PDG2020} 
    & $\pm 0.002 $ (68$\%$ CL)~\cite{PDG2020} &
    $< 3.0$ (90$\%$ CL)\\
    $10^{10}\times$B($K^+ \to \pi^+ \nu \bar{\nu}$) & $1.7 \pm 1.1$ (68$\%$ CL)~\cite{PDG2020} 
    & $\pm 0.04 $ (68$\%$ CL)~\cite{PDG2020} &
    $1.7^{+2.16}_{-1.70}$\\
    $\Delta M_{B_s}$ [ps$^{-1}$] & $17.757 \pm 0.021$ (68$\%$ CL)~\cite{HFAG2019, PDG2020} 
    & $\pm 2.7$ (68$\%$ CL)~\cite{DeltaMBs_SM} &
    $17.757 \pm 5.29$\\
    $10^4\times$B($b \to s \gamma)$ & $3.32 \pm 0.15$ (68$\%$ CL)~\cite{HFAG2019, PDG2020} 
    & $\pm 0.23$ (68$\%$ CL)~\cite{Misiak_2015} &  $3.32\pm 0.54$\\
    $10^6\times$B($b \to s~l^+ l^-$)& $1.60 ~ ^{+0.48}_{-0.45}$ (68$\%$ CL)~\cite{bsll_BABAR_2014}
    & $\pm 0.11$ (68$\%$ CL)~\cite{Huber_2008} & $1.60 ~ ^{+0.97}_{-0.91}$\\
    $(l=e~{\rm or}~\mu)$ &&&\\
    $10^9\times$B($B_s\to \mu^+\mu^-$) & $2.69~^{+0.37}_{-0.35}$ (68$\%$CL)~\cite{B@ICHEP2020}
    & $\pm0.23$  (68$\%$ CL)~\cite{Bsmumu_SM_Bobeth_2014} 
    & $2.69~^{+0.85}_{-0.82}$ \\
    $10^4\times$B($B^+ \to \tau^+ \nu $) & $1.06 \pm 0.19$ (68$\%$CL)
    ~\cite{HFAG2019}
    &$\pm0.29$  (68$\%$ CL)~\cite{Btotaunu_LP2013} & $1.06 \pm 0.69$\\
    $ m_{h^0}$ [GeV] & $125.09 \pm 0.24~(68\%~ \rm{CL})$ \cite{Higgs_mass_ATLAS_CMS}
    & $\pm 3$~\cite{Higgs_mass_Heinemeyer} & $125.09 \pm 3.48$ \\
    $\kappa_b$ & $1.06^{+0.37}_{-0.35}~(95\%~ \rm{CL})$ \cite{kappa_bgamg_ATLAS}
    &  & $1.06^{+0.37}_{-0.35}$ (ATLAS)\\
    & $1.17^{+0.53}_{-0.61}~(95\%~ \rm{CL})$ \cite{kappa_bgamg_CMS}
    &  & $1.17^{+0.53}_{-0.61}$ (CMS)\\
    $\kappa_g$ & $1.03^{+0.14}_{-0.12}~(95\%~ \rm{CL})$ \cite{kappa_bgamg_ATLAS}
    &  & $1.03^{+0.14}_{-0.12}$ (ATLAS)\\
    & $1.18^{+0.31}_{-0.27}~(95\%~ \rm{CL})$ \cite{kappa_bgamg_CMS}
    &  & $1.18^{+0.31}_{-0.27}$ (CMS)\\
    $\kappa_\gamma$ & $1.00 \pm 0.12~(95\%~ \rm{CL})$ \cite{kappa_bgamg_ATLAS}
    &  & $1.00 \pm 0.12$ (ATLAS)\\
    & $1.07^{+0.27}_{-0.29}~(95\%~ \rm{CL})$ \cite{kappa_bgamg_CMS}
    &  & $1.07^{+0.27}_{-0.29}$ (CMS)\\
&&&\\
    \hline
\end{tabular}
\end{center}
\label{TabConstraints}}
\end{table*}
%
\hspace{1cm}

\begin{itemize}

\item
The LHC limits on sparticle masses (at 95\% CL)~\cite{SUSY@LP2019, SUSY@ICHEP2020, 
SUSY_summary_plot@ATLAS, SUSY_summary_plot@CMS, SUSY@ATLAS2020}:

We impose conservative limits for safety though actual limits are 
somewhat weaker than those shown here.
In the context of simplified models, gluino masses $\msg \lesssim 2.35~{\rm TeV}$ are 
excluded for $\mnt1 < 1.55~{\rm TeV}$. There is no gluino mass limit for $\mnt1 > 1.55~{\rm TeV}$.
The 8-fold degenerate first two generation squark masses are excluded below 1.92~TeV for $\mnt1 < 0.9~{\rm TeV}$.
There is no limit on the masses for $\mnt1 > 0.9~{\rm TeV}$. 
We impose this squark mass limit on $m_{\su_3}$ and $m_{\sd_3}$. 
Bottom-squark masses are excluded below 1.26~TeV for $\mnt1 < 0.73~{\rm TeV}$. 
There is no bottom-squark mass limit for $\mnt1 > 0.73~{\rm TeV}$.
Here the bottom-squark mass means the lighter sbottom mass $m_{\sb_1}$.
We impose this limit on $m_{\sd_1}$ since $\sd_1 \sim \sb_R$ (see Table \ref{flavourdecomp}).
A typical top-squark mass lower limit is $\sim$ 1.26~TeV for $m_{\nt_1} < 0.62$ TeV. 
There is no top-squark mass limit for $m_{\nt_1} > 0.62$ TeV. 
Here the top-squark mass means the lighter stop mass $m_{\st_1}$.
We impose this limit on $m_{\su_1}$ since $\su_1 \sim \st_R$ (see Table \ref{flavourdecomp}).
For sleptons/sneutrinos heavier than the lighter chargino $\ch_1$ and the second neutralino $\nt_2$, 
the mass limits are $m_{\ch_1}, m_{\nt_2} > 0.74$ TeV for $m_{\nt_1} \lesssim 0.3$ TeV and 
there is no $m_{\ch_1}$, $m_{\nt_2}$ limits for $m_{\nt_1} > 0.3$ TeV; 
For sleptons/sneutrinos lighter than $\ch_1$ and $\nt_2$, 
the mass limits are $m_{\ch_1}, m_{\nt_2} > 1.15$ TeV for $m_{\nt_1} \lesssim 0.72$ TeV and 
there is no $m_{\ch_1}$, $m_{\nt_2}$ limits for $m_{\nt_1} > 0.72$ TeV.
For mass degenerate selectrons $\ti{e}_{L,R}$ and smuons $\ti{\mu}_{L,R}$, masses below 
0.7 TeV are excluded for $m_{\nt_1} < 0.41$ TeV. For mass degenerate staus $\stau_L$ and 
$\stau_R$, masses below 0.39 TeV are excluded for $m_{\nt_1} < 0.14$ TeV. 
There is no sneutrino $\ti{\nu}$ mass limit from LHC yet.
Sneutrino masses below 94 GeV are excluded by LEP200 experiment \cite{PDG2020}.

\item
The constraint on ($m_{A^0, H^+}, \tan\beta$) (at 95\% CL) from searches for the MSSM Higgs bosons 
$H^0$, $A^0$ and $H^+$ at LHC,~\cite{H_to_tautau@ATLAS,H_to_tautau@CMS,H_tb@ATLAS,H_tb@ATLAS_2021,H_tau_nu@ATLAS,H_tb@CMS,H_tau_nu@CMS}, 
where $H^0$ is the heavier $CP$-even Higgs boson.

\item
The experimental limit on SUSY contributions on the electroweak
$\rho$ parameter ~\cite{Altarelli:1997et}: $\Delta \rho~ (\rm SUSY) < 0.0012.$

\end{itemize}



Furthermore, we impose the following theoretical constraints from the vacuum 
stability conditions for the trilinear coupling matrices~\cite{Casas}: 
\begin{eqnarray}
|T_{U\alpha\alpha}|^2 &<&
3~Y^2_{U\alpha}~(M^2_{Q \alpha\alpha}+M^2_{U\alpha\alpha}+m^2_2)~,
\label{eq:CCBfcU}\\[2mm]
|T_{D\alpha\alpha}|^2 &<&
3~Y^2_{D\alpha}~(M^2_{Q\alpha\alpha}+M^2_{D\alpha\alpha}+m^2_1)~,
\label{eq:CCBfcD}\\[2mm]
|T_{U\alpha\beta}|^2 &<&
Y^2_{U\gamma}~(M^2_{Q \beta\beta}+M^2_{U\alpha\alpha}+m^2_2)~,
\label{eq:CCBfvU}\\[2mm]
|T_{D\alpha\beta}|^2 &<&
Y^2_{D\gamma}~(M^2_{Q \beta\beta}+M^2_{D\alpha\alpha}+m^2_1)~,
\label{eq:CCBfvD}
\end{eqnarray}
where
$\a,\b=1,2,3,~\a\neq\b;~\gamma={\rm Max}(\a,\b)$ and
$m^2_1=(m^2_{H^+}+m^2_Z\sin^2\theta_W)\sin^2\b-\frac{1}{2}m_Z^2$,
$m^2_2=(m^2_{H^+}+$  
$m^2_Z\sin^2\theta_W)$ $\cos^2\beta-\frac{1}{2}m_Z^2$.
The Yukawa couplings of the up-type and down-type quarks are
$Y_{U\alpha}=\sqrt{2}m_{u_\alpha}/v_2=\frac{g}{\sqrt{2}}\frac{m_{u_\alpha}}{m_W
\sin\beta}$
$(u_\a=u,c,t)$ and
$Y_{D\alpha}=\sqrt{2}m_{d_\alpha}/v_1=\frac{g}{\sqrt{2}}\frac{m_{d_\alpha}}{m_W
\cos\beta}$
$(d_\a=d,s,b)$,
with $m_{u_\a}$ and $m_{d_\a}$ being the running quark masses at the 
scale $\rm Q=1$~TeV and $g$ being the SU(2) gauge coupling. All soft SUSY-breaking parameters 
are given at $\rm Q=1$~TeV. As SM parameters we take $m_Z=91.2~\gev$ and
the on-shell top-quark mass $m_t=172.9~\gev$ \cite{PDG2020}.
\end{appendix}
%

%

\end{document}